# JWST observations of the Horsehead photon-dominated region I. First results from multi-band near- and mid-infrared imaging

A. Abergel[1], K. Misselt[2], K.D. Gordon[3], A. Noriega-Crespo[3], P. Guillard[4], D. Van De Putte[3], A.N. Witt[5], N. Ysard[6, 1], M. Baes[7], H. Beuther[8], P. Bouchet[9], B.R. Brandl[10], M. Elyajouri[1], O. Kannavou[1], S. Kendrew[12], P. Klassen[13], and B. Trahin[3]

*(Affiliations can be found after the references)*

**ABSTRACT**

*Context.* The *James Webb* Space Telescope (JWST) has captured the sharpest infrared images ever taken of the Horsehead nebula, a prototypical moderately irradiated photon-dominated region (PDR) that is fully representative of most of the UV-illuminated molecular gas in the Milky Way and star-forming galaxies.

*Aims.* We investigate the impact of far-ultraviolet (FUV) radiation emitted by a massive star on the edge of a molecular cloud in terms of photoevaporation, ionization, dissociation, $H_2$ excitation, and dust heating. We also aim to constrain the structure of the edge of the PDR and its illumination conditions.

*Methods.* We used NIRCam and MIRI to obtain 17 broadband and 6 narrowband maps of the illuminated edge of the Horsehead across a wide spectral range from 0.7 to 28 $\mu$m. We mapped the dust emission, including the aromatic and aliphatic infrared (IR) bands, scattered light, and several gas phase lines (e.g., Pa$\alpha$, Br$\alpha$, $H_2$ 1-0 S(1) at 2.12 $\mu$m). For our analysis, we also associated two HST-WFC3 maps at 1.1 and 1.6 $\mu$m, along with HST-STIS spectroscopic observations of the H$\alpha$ line.

*Results.* We probed the structure of the edge of the Horsehead and resolved its spatial complexity with an angular resolution of 0.1 to 1″ (equivalent to $2 \times 10^{-4}$ to $2 \times 10^{-3}$ pc or 40 to 400 au at the distance of 400 pc). We detected a network of faint striated features extending perpendicularly to the PDR front into the H II region in NIRCam and MIRI filters sensitive to nano-grain emission, as well as in the HST filter at 1.1 $\mu$m, which traces light scattered by larger grains. This may indeed figure as the first detection of the entrainment of dust particles in the evaporative flow. The filamentary structure of the 1-0 S(1) line of $H_2$ at the illuminated edge of the PDR presents numerous sharp sub-structures on scales as small as 1.5″. An excess of $H_2$ emission compared to dust emission is found all along the edge, in a narrow layer (width around 1″, corresponding to $2 \times 10^{-3}$ pc or 400 au) directly illuminated by $\sigma$-Orionis. The ionization front and the dissociation front appear at distances 1 − 2″ behind the external edge of the PDR and seem to spatially coincide, indicating a very small thickness of the neutral atomic layer (below 100 au). All broadband maps present strong color variations between the illuminated edge and the internal regions. This can be explained by dust attenuation in a scenario where the illuminating star $\sigma$-Orionis is slightly inclined compared to the plane of the sky, so that the Horsehead is illuminated from behind at an oblique angle. The deviations from predictions of the measured emissions in the H$\alpha$, Pa$\alpha$, and Br$\alpha$ lines also indicate dust attenuation. With a very simple model, we used the data to derive the main spectral features of the extinction curve. A small excess of extinction at 3 $\mu$m may be attributed to icy $H_2O$ mantles onto grains formed in dense regions. We also derived attenuation profiles from 0.7 to 25 $\mu$m across the PDR. In all lines of sight crossing the inner regions of the Horsehead, especially around the IR peak position, it appears that dust attenuation is non-negligible over the entire spectral range of the JWST.

## 1. Introduction

Photon-dominated regions (PDRs) are predominantly neutral regions of the interstellar medium (ISM) in which the heating and chemistry are mainly regulated by far ultraviolet (FUV) photons from massive stars (e.g., Hollenbach & Tielens 1999; Wolfire et al. 2022). These regions form at the broad interface between young stars and molecular clouds, where UV radiation heats gas and dust and drives chemical evolution. The interaction of stellar radiation with in situ material includes: (1) dust processing, including the disruption of grain mantles and clusters as well as larger coagulated grains formed in shielded dense regions; (2) dust heating; and (3) ionization, dissociation, and heating of the atomic and molecular gas. These processes can be strongly stratified in the PDR boundary layers (ionized→ neutral→ molecular gas) due to different species absorbing UV photons at different energies and the attenuation of the UV radiation as we go deeper into the PDR. They are active on scales of hundreds of au (on the order of ≤ 1″ at the typical distances of the most studied Galactic PDRs). This indicates that the physical conditions vary dramatically on small spatial scales, dependent on both the local structure of the molecular cloud as well as the input radiation field (dependent on the number, size, distance, and spectral types of the surrounding stars undergoing excitation).

Molecule formation (CO, $H_2$, etc.) and chemistry, grain growth, and destruction deeper in the molecular cloud are governed by the energetic and chemical input at the PDR boundary. Therefore, much of the molecular and atomic gas and dust in the Galaxy resides in PDRs and the bulk of the continuum and line emission observed from both gas and dust is seen to emerge from PDR regions. As PDRs represent the interface between sites of active star formation and the more dense molecular material that stars end up forming out of, analyses of the detailed energetic and chemical processes that occur within them provide insights into the environmental and feedback conditions that dictate star formation.

The physical and chemical processes in PDRs – dust grain and molecular formation, growth, and destruction, atomic and molecular heating, cooling, and excitation among them – produce observational signatures across the IR wavelength regime. These observational signatures include numerous atomic and molecular lines, aromatic and aliphatic bands, dust continuum, and extended red emission (ERE) at the shortest wavelengths.

The importance of the PDR environment-shaping chemistry and physics of the ISM and star formation at small scales, as well as their role in shaping the global emission from a galaxy, has prompted a significant investment on the part of *James Webb* Space Telescope (JWST) operations into PDR studies. These investments include the Early Release Science (ERS) program on the Orion Bar (Berné et al. 2022; Habart et al. 2023; Peeters et al. 2023) as well as the Guaranteed Time Observations (GTO) program presented here. This GTO program (PID 01192) is a collaboration between the MIRI-US, MIRI-EC, and NIRCam teams, with a combined time investment of 45.2 hours, focusing on two well-known PDRs, NGC 7023 and the Horsehead Nebula. These objects were selected for their proximity, around 400 pc (Anthony-Twarog 1982; Gaia Collaboration et al. 2018), relatively simple geometry, and excitation conditions; namely, $G_0 \sim 100$ and 1500 for the Horsehead and NGC 7023, respectively in units of the Habing (1968) field, which is significantly smaller than the atypical value of $\sim 5 \times 10^4$ for the Orion Bar. Their proximity, combined with the high spatial resolution of JWST, provides milli-pc resolution of the physics and chemistry of the PDR. In addition, their nearly edge-on orientation to the line of sight simplifies the interpretation of the evolution of the gas, dust, and molecular populations across the PDR front.

Given the broad wavelength range where gas and dust signatures manifest and their disparate characteristics across PDRs, we elected to employ NIRSpec, NIRCam, and MIRI on JWST to achieve a wavelength coverage from roughly 0.7 $\mu$m to 28 $\mu$m with both imaging (NIRCam + MIRI imagers) and integral field unit (IFU) spectroscopy (NIRSpec + MIRI). Here, we focus on NIRCam and MIRI imaging data for the Horsehead nebula; the Horsehead MIRI and NIRSpec IFU data are presented in a companion paper. These data were obtained in JWST Cycle 1 while NIRSpec IFU and all NGC 7023 observations were deferred to Cycle 2.

The paper is organized as follows. In Sect. 2, we describe the main physical characteristics of the Horsehead nebula inferred from previous studies. In Sect. 3, we detail the NIRCam and MIRI observations and data reduction, the ancillary HST maps used, and the contributions of lines, bands, and continuum emission in all selected filters. In Sects. 4 and 5, we analyze our set of images for the ionized region facing the PDR and for the illuminated edge of the PDR, respectively. New constraints placed on the physical structure of the PDR are discussed in Sect. 6. The analysis of of dust emission and scattering across the illuminated edge is presented in Sect. 7. In Sect. 8, we offer a quantitative study of the dust attenuation across the PDR. In Sect. 9, we present our conclusions.

## 2. Horsehead nebula

The Horsehead is located at the Western side of the Orion B molecular cloud which is situated at distance ∼ 400 pc (Anthony-Twarog 1982) from the Sun. In front of the western illuminated edge of Orion B, the images taken in the visible spectral range are dominated by diffuse emission due to H$\alpha$ emission line emerging from the H II region IC 434 and the Horsehead emerges from this edge as a dark cloud seen in silhouette against the background plasma in IC 434 (e.g., de Boer 1983; Neckel & Sarcander 1985; Compiègne et al. 2007; Bally et al. 2018).

The Horsehead is powered by the O9.5V binary system $\sigma$-Orionis (Warren & Hesser 1977) with an effective temperature of $T_{\rm eff} \sim 34600$ K (Schaerer & de Koter 1997), located at a projected distance $d_{\rm edge} \sim 3.5$ pc from the edge of the Horsehead PDR. At this projected separation, the far-UV (FUV) intensity of the incident radiation field at the edge can be estimated to be $G_0 \sim 100$ (e.g., Habart et al. 2005). As the Horsehead is seen predominantly in silhouette against IC 434, it is likely mostly illuminated on its backside by $\sigma$-Orionis (Pound et al. 2003). It is undergoing evaporation due to the UV radiation field, with an estimated survival time of $5 \times 10^6$ years (Pound et al. 2003).

The Horsehead is archetypical of moderately excited PDRs. Such PDRs are much more common than highly excited PDRs like the Orion Bar (Hughes et al. 2015) and are representative of most of the UV-illuminated molecular gas in the Milky Way and star-forming galaxies. Moreover, thanks to its proximity and its nearly edge-on geometry, the Horsehead is an ideal target for studying the physical structures of PDRs and the evolution of the physico-chemical characteristics of the gas and dust with the local conditions (density and illumination).

The first mid-IR observations of the Horsehead were conducted with the ISOCAM imager of ISO (Abergel et al. 2002, 2003), then the IRAC and MIPS cameras onboard Spitzer (Schirmer et al. 2020). Because of the low-excitation conditions, the mid-IR emission is mainly due to the stochastic heating of nano-grains (size below ∼ 20 nm) (e.g., Sellgren et al. 1983; Boulanger et al. 1998), including interstellar polycyclic aromatic hydrocarbons (e.g., Leger & Puget 1984; Weingartner & Draine 2001; Jones et al. 2013). Therefore the mid-IR emission is a direct tracer of the UV radiation field and the density. The ISO and Spitzer maps reveal that the illuminated edge of the PDR is delineated by bright and narrow mid-IR filaments due to the combined effect of density increase on the illuminated side and extinction of the incident radiation inside the dense cloud (with gas density a few times $10^4$ cm$^{-3}$). These IR filaments have a total width of 10″ to 20″, depending on the position along the edge and presents substructures that are not resolved by ISO and Spitzer. The illuminated edge of the PDR is also delineated by bright and narrow filaments in the 1-0 S(1) line of $H_2$ at 2.12 $\mu$m (Habart et al. 2005). As this line is very sensitive to both the FUV radiation field and the gas density, the comparison of the observations with PDR model calculations has demonstrated that the gas density follows a steep density gradient at the illuminated edge, with a scale length of 0.02 pc (or ∼10″) and $n \sim 10^4$ cm$^{-3}$ and $n \sim 10^5$ cm$^{-3}$ in the $H_2$ emitting and inner molecular layers, respectively. Habart et al. (2005) have also shown that the Horsehead is not viewed completely edge-on, but with a small inclination estimated to be ∼ 6°. We are therefore seeing a fraction of the surface of the cloud in the observations.

The far-infrared (FIR) continuum emission of the Horsehead has been mapped with Herschel at low angular resolution (Schirmer et al. 2020). This emission is due to large grains in thermal equilibrium with the incident radiation field, with amorphous silicate and carbon likely with core-mantle and aggregate structure (e.g., Ormel et al. 2011; Jones et al. 2013). The FIR maps also exhibit bright filaments at the illuminated edge, due to the combined effects of the increase of density at the edge and the decrease of the radiation field in the dense regions.

The proximity of the Horshehead and its unique geometry have also encouraged a number of studies at millimeter wave-lengths, with increasing angular resolutions. Pound et al. (2003) have used CO $J = 1-0$ observations to derive the total mass (27 $M_\odot$) and the mean density ($5 \times 10^3$ cm$^{-3}$) of the Horsehead. $C^{18}O$ $J = 2-1$ observations have allowed Hily-Blant et al. (2005) the study of the inner layers in the Horsehead. The molecular content behind the illuminated edge of the PDR has been extensively studied with the IRAM 30 m and PdBI interferometers (Goicoechea et al. 2006; Pety et al. 2007; Gerin et al. 2009; Goicoechea et al. 2009, etc.). Several complex organics such as $H_2CO$, $CH_3OH$, $CH_3CN$, $CH_2CO$, and $CH_3CHO$ were also found (Guzmán et al. 2011, 2013; Gratier et al. 2013; Guzmán et al. 2015), suggesting that these species are producted via grain surface reactions in the UV-dominated PDR (Le Gal et al. 2017). The kinematics of the Horsehead and the IC 434 H II regions has been analysed by Bally et al. (2018) using CO and $C^+$ observations. Finally, the observations of the millimetric molecular emission with the highest angular resolution ($\sim 0.5''$) recently taken at ALMA have enabled Hernández-Vera et al. (2023) to highligh a very sharp transition (< 650 au) between the molecular and ionized gas at the edge of the PDR.

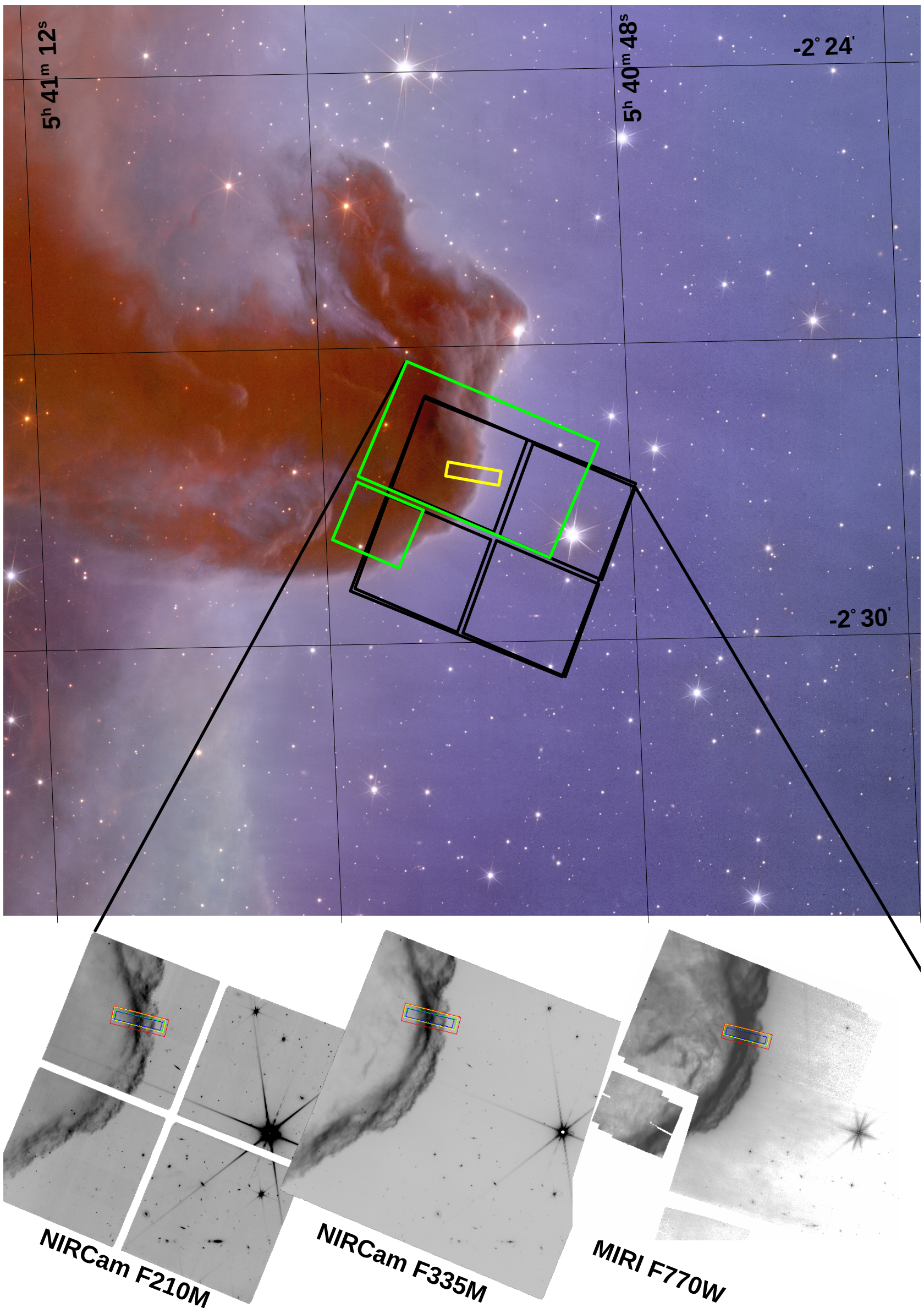

Fig. 1: FOV of imagers on the sky. The color image is from the recent Euclid ERO release (ESA/Euclid/Euclid Consortium/NASA, image processing by J.-C. Cuillandre (CEA Paris-Saclay), G.Anselmi) combining maps taken by the VIS imager (0.55-0.9 $\mu$m) and the NISP spectrophotometer in the Y (0.99-1.192 $\mu$m) and H bands (0.92 -1.25 $\mu$m). The footprint of all NIRCam images is shown in black; the MIRI imaging footprint for all filters is shown in green; and the location of the IFU (MIRI and NIRSpec) footprints in yellow. Representative images for MIRI and the NIRCam short and long wave channels are shown below. In these JWST images, the observed locations of the MIRI IFU channel 1, 2, 3, and 4 footprints are shown in blue, green, yellow, and red, respectively. All NIRCam SW mosaics consist of 4 individual detectors, B1, B2, B4, and B3, clockwise from the top; the bulk of the PDR front falls on NIRCam detector B1.

Table 1: Summary of data

| **Filter**[1] | $\lambda_{pivot}$ [$\mu$m] | Bandwidth [ $\mu$m ] | **NGROUP** | $T_{int}$[2] [sec] | **S/N**[3] | **Note**[4] |
|---|---|---|---|---|---|---|
| | | | **NIRCam - OBSID 015, All SHALLOW4** | | | |
| | | | **Short-wave Channel** | | | |
| F070W | 0.704 | 0.128 | 7 | 1095.15 | 22, 12, 10 | H$\alpha$+ Continuum + ERE |
| F090W | 0.901 | 0.194 | 7 | 1095.15 | 19, 16, 14 | Continuum + ERE |
| F140M | 1.401 | 0.142 | 7 | 1095.15 | 18, 22, 14 | Continuum |
| F164N | 1.644 | 0.020 | 10 | 1578.31 | 3, 3, 2 | [Fe II] |
| F187N | 1.874 | 0.024 | 10 | 1578.31 | 18, 13, 6 | HI (Pa$\alpha$) |
| F210M | 2.093 | 0.205 | 7 | 1095.15 | 19, 33, 15 | Continuum |
| F212N | 2.120 | 0.027 | 10 | 1578.31 | 10, 19, 6 | $H_2$ [1-0 S(1)] |
| | | | **Long-wave Channel** | | | |
| F250M | 2.503 | 0.181 | 7 | 1095.15 | 32, 115, 29 | Continuum |
| F300M | 2.996 | 0.318 | 7 | 1095.15 | 44, 99, 25 | Continuum |
| F335M | 3.365 | 0.347 | 7 | 1095.15 | 129, 410, 61 | AIB C-H stretch + Aliph. band + Continuum |
| F323N | 3.237 | 0.038 | 10 | 1578.31 | 24, 68, 9 | $H_2$ [1-0 O(5)] |
| F405N | 4.055 | 0.046 | 10 | 1578.31 | 28, 37, 13 | H1 (Br$\alpha$) |
| F430M | 4.280 | 0.228 | 7 | 1095.15 | 37, 87, 24 | $CO_2$ + Continuum |
| F470N | 4.707 | 0.051 | 10 | 1578.31 | 18, 41, 10 | $H_2$ [(0-0 S(9)] |
| | | | **MIRI - OBSID 014, All FASTR1** | | | |
| F560W | 5.6 | 1.2 | 12 | 99.901 | 31,107,15 | Continuum |
| F770W | 7.7 | 2.2 | 12 | 99.901 | 134, 519, 69 | AIB, C-C stretch + Continuum |
| F1000W | 10.0 | 2.0 | 12 | 99.901 | 53, 184, 16 | Continuum |
| F1130W | 11.3 | 0.7 | 12 | 99.901 | 54, 179, 15 | AIB, C-H out of plane + Continuum |
| F1280W | 12.8 | 2.4 | 12 | 99.901 | 84, 265, 26 | Continuum |
| F1500W | 15.0 | 3.0 | 12 | 99.901 | 52, 193, 24 | Continuum |
| F1800W | 180.0 | 3.0 | 12 | 99.901 | 41, 141, 14 | Continuum |
| F2100W | 21.0 | 5.0 | 12 | 99.901 | 29, 107, 12 | Continuum |
| F2550W | 25.5 | 4.0 | 12 | 99.901 | 17, 47, 11 | Continuum |
| | | | **HST/WFC3** | | | |
| F110W | 1.16 | 0.5 | | 1106 | 75, 70, 53 | Continuum |
| F160W | 1.54 | 0.3 | | 1406 | 70, 82, 54 | Continuum |
| | | | **HST/STIS**[5] | | | |
| G750L, 52″ × 2″ | 0.52-1.0 | 0.016 | | 4496 | 10, 12, 3 | H$\alpha$ |

[1] NIRCam SW and LW are observed simultaneously, so a single observation/configuration provides 2 filter measurements.
[2] Total time in the filter over 3 dither positions.
[3] S/N was estimated at three points spanning the PDR front in the region covered by the MIRI and NIRSpec IFUs (Fig. 1); just outside the front, at the peak of the front, and into the molecular region.
[4] $H_2$ transitions presented as $v_u$–$v_l$ $\Delta J(J_l)$
[5] Only partial STIS data were available at time of publication. We give the grating and slit size, and rather than $\lambda_{pivot}$ and bandwidth, we report the wavelength coverage and approximate extended source resolution, respectively.

## 3. Data

Here we provide a brief description of the data obtained in this program[1]. The orientation of the MIRI, NIRCam, and IFU (MIRI/MRS and NIRSpec) footprints with respect to the larger Horsehead complex is shown in Fig. 1, in black, green and yellow, respectively. Due to the limited field of view of the MIRI and NIRSpec IFUs, we restricted the IFU maps to narrow mosaics along the line of sight from the exciting star, perpendicular to the PDR front and centered roughly on regions covered in previous datasets (e.g., Abergel et al. 2003; Schirmer et al. 2020), with overlapping wide-field imaging. The imaging program was designed to center the location of the IFU footprint (in yellow on the top panel of Fig. 1) on a single detector of the imaging instruments. We estimated exposure times from the ETC[2] using the Orion bar template SED described in Berné et al. (2022) scaled to the surface brightness of the Horsehead PDR front derived from Spitzer imaging and targeting an signal-to-noise ratio (S/N) per a resolution element of 10 in the narrowband NIRCam SW filters at the peak of the PDR front. Given the simultaneous SW/LW NIRCam imaging, the requisite exposure time in the SW filters resulted in exceeding the per resolution element target S/N in all the LW filters (larger pixels, generally wider bandpasses). In the case of MIRI, the exposure time was set largely by a reasonable minimal ramp length in the FASTR1 mode. In all MIRI filters, the S/N exceeds 10, in some cases significantly (see Table 1).

[1] The data can be downloaded from the MAST portal https://mast.stsci.edu/portal_jwst/Mashup/Clients/Mast/Portal.html

[2] https://jwst.etc.stsci.edu/

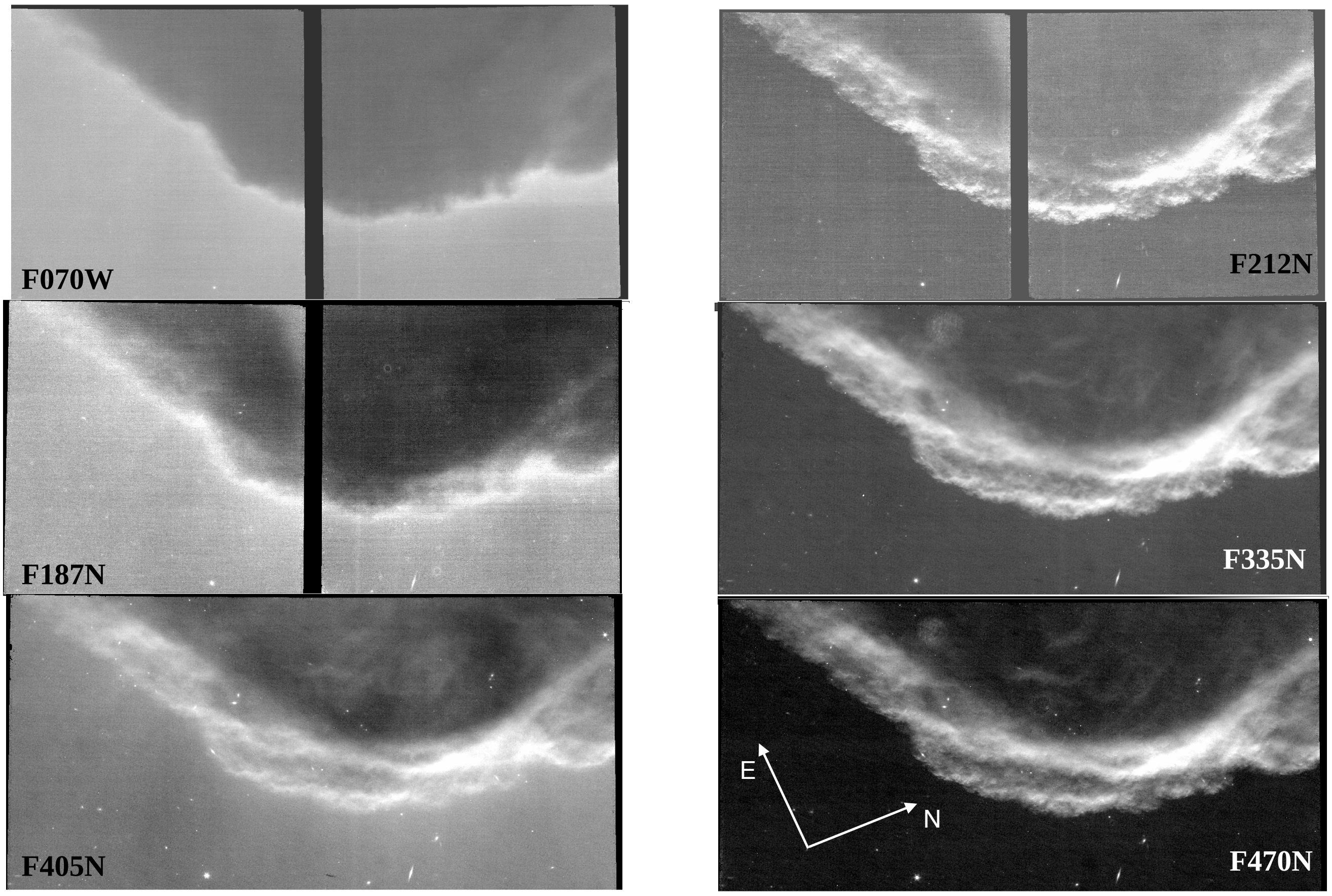

Fig. 2: PDR front in six NIRCam filters sensitive to ionized (left) and molecular (right) hydrogen. Left panel, from top to bottom: F070W (H$\alpha$), F187N (Pa$\alpha$), and F405N (Br$\alpha$). Right panel, from top to bottom: F212N [1-0 S(1)], F323N [1-0 O(5)], and F470N [0-0 S(9)]. Logarithmic scales are used for all maps. Note: we do not show the F164N filter; designed to be sensitive to [Fe II], almost no signal above the instrumental noise and artifacts was detected.

### 3.1. NIRCam

NIRCam[3] (Rieke et al. 2023) is a near-IR (0.7-5 $\mu$m) instrument. It is comprised of two (A and B) fully redundant imaging modules. Each module images spatially matched short (SW - 0.6-2.3 $\mu$m) and long (LW - 2.4-5.0 $\mu$m) wavelength channels simultaneously via a dichroic beam splitter. Each SW module consists of 4 detectors with a pixel scale of ~0.03 ″/pixel while the LW module consists of a single detector with a pixel scale of ~0.06 ″/pixel. For this program, we utilize 14 science imaging filters (7 in each of the SW and LW channels) designed to cover H recombination and $H_2$ lines, aromatic/aliphatic features, and the continuum, in the NIRCam wavelength coverage. Given that the short wave module on NIRCam is comprised of four individual detectors with gaps of $\sim 5''$ between them, we designed the NIRCam imaging program so that the region of IFU coverage across the front was centered on a single SW detector; given the simultaneous and spatially matched LW imaging, this centers the PDR front in a quadrant of the LW detector.

NIRCam supports multiple readout modes to balance data volume against sensitivity to cosmic rays. All our NIRCam data utilized the SHALLOW4 mode in which four samples are co-added on board, and a single readout frame is dropped. Based on ETC predictions, we specified ten SHALLOW4 groups in the narrowband filters and seven in the medium and wide band filters with three dither positions, resulting in total on-source exposure times of ~1600 and 1100 seconds, respectively. While the co-adds in SHALLOW4 attenuate the sensitivity to cosmic rays, they provide a significant gain in read noise, especially for the narrowband filters. During the NIRCam imaging, we enabled parallel MIRI imaging; this ancillary MIRI imaging off the PDR was used as a check on astronomical and instrumental background for the MIRI science imaging (§3.2). The details of the NIRCam data are summarized in Table 1.

The NIRCam data were analyzed with the STScI pipeline[4] as well as using custom software developed during instrument design and testing. For the data presented here, we elected to utilize a custom reduction for Level1B processing (a.k.a. ramps-to-slopes); our custom approach implements an inter-pixel capacitance correction by default as well as avoiding propagation of bad pixels to the eight nearest neighbors via the adaptive saturation step in the default pipeline[5]. The output of our custom ramp fitting step was fed to the Level 2 pipeline for calibration, and the coordinates were updated using the GAIA DR3 catalog (Gaia Collaboration et al. 2023) prior to processing by the Level 3 pipeline for mosaicing. All NIRCam images can be seen in Figs. 2 and 3.

[3] https://jwst-docs.stsci.edu/jwst-near-infrared-camera

[4] https://jwst-docs.stsci.edu/jwst-science-calibration-pipeline-overview

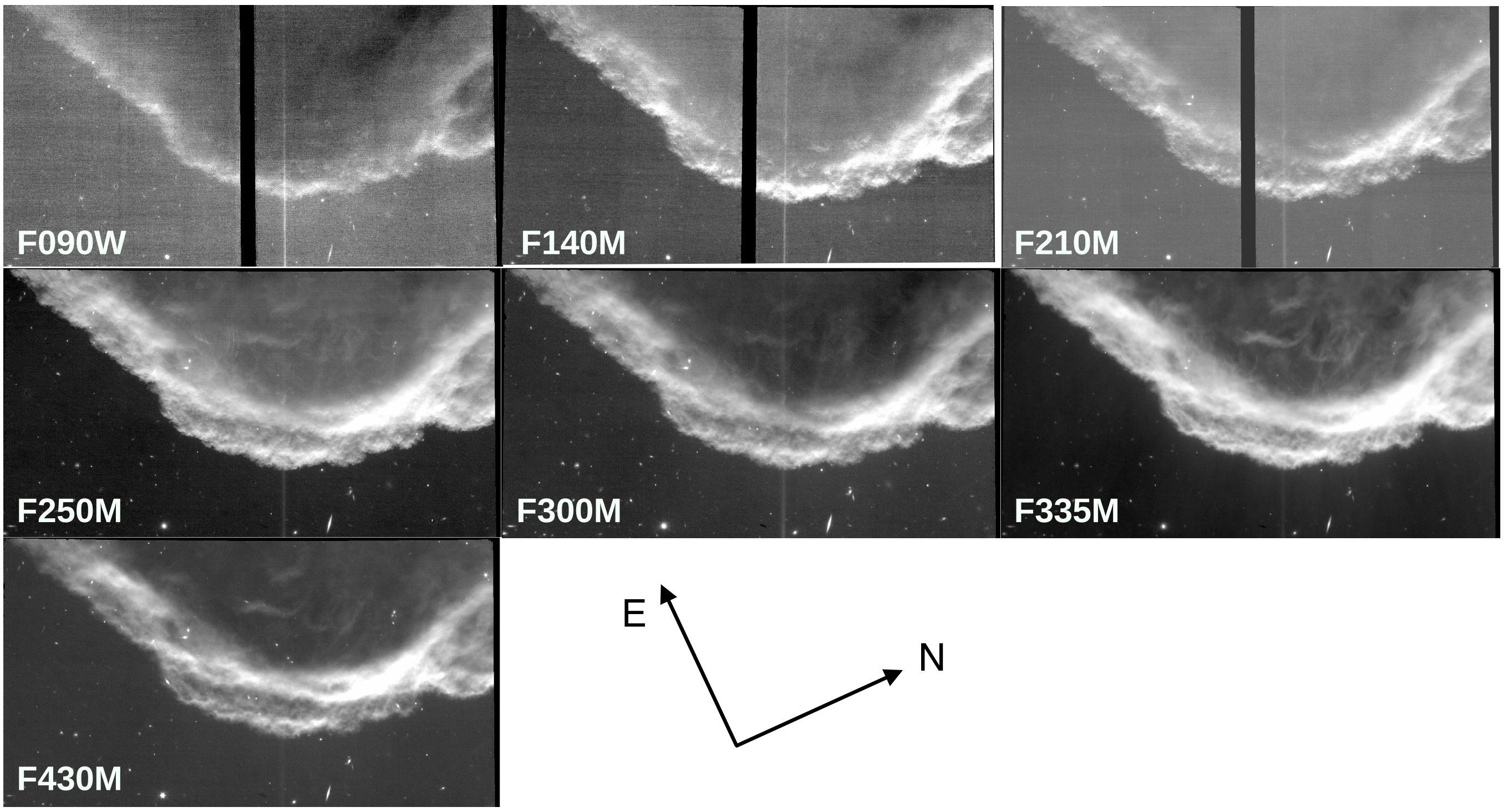

Fig. 3: PDR front in remaining seven continuum/PAH NIRCam filters. Left to right, from top to bottom: F090W, F140M, F210M, F250M, F300M, F335M, and F430M. Logarithmic scales are used for all maps.

The absolute flux calibration uncertainties for point sources are on the order of $< 1-2\%$ for the NIRCam filters utilized here (Boyer et al. in prep). As we are dealing with extended sources, we double the point source absolute flux calibration errors and conservatively estimate the extended source calibration uncertainty to be $\sim 4\%$ for all filters.

Poorly understood noise sources in the NIR JWST detectors commonly fall under the rubric of "$1/f$" noise. This $1/f$ noise manifests in the fast readout direction - in the case of NIRCam, the row readout direction - and likely has its origin in common reference voltages in the readout electronics. In processed images, $1/f$ noise has a characteristic row stripping appearance. Many of the standard post-processing $1/f$ corrections (e.g., Schlawin et al. 2020) rely on source-free regions in the fast (row) read direction. Given that our source is extended and covers many full fast-read rows, such algorithms are not appropriate for our data set. However, $1/f$ corrections that model the background can give a better representation of the $1/f$ noise component in the case of extended sources like ours. To that end, we employed an algorithm developed by Chris Willot[6] to mitigate $1/f$ noise stripping in our data. We found significant improvement in the $1/f$ noise "contamination" and while an examination of the correction residuals revealed some systematic residuals associated with the extended structure, the residuals in the source locations were less than 1%.

### 3.2. MIRI

MIRI[7] (Wright et al. 2023) is the mid-infrared (MIR: 5-28 $\mu$m) imager and spectrometer on JWST. MIRI's capabilities include imaging (direct and coronographic), low-resolution ($\sim$100) slit or slitless spectroscopy (5-14 $\mu$m), and medium-resolution ($\sim$ 1500-3000) IFU spectroscopy (MRS) across the full wavelength range. In direct imaging, the MIRI instrument supplies nine wide band filters (1-5 $\mu$m bandwidth; see Table 1). For this program, we utilized the direct imaging capabilities to achieve broad spatial coverage of the region along with the mapping capabilities of the MRS to obtain spectra across the PDR front. Here we focus on the imaging program; for details of the MRS IFU observations, see Guillard et al. in prep. MIRI provides a single imaging detector with a pixel scale of ~0.11″/pixel, and the center of the PDR front was simply placed at the center of the detector's field of view (FOV). For this program, we utilized all nine available imaging filters in order to map the aromatic features and the continuum in the MIRI wavelength coverage.

[5] This issue has been mitigated in the pipeline by updates to reference files.

[6] https://github.com/chriswillott/jwst

[7] https://jwst-docs.stsci.edu/jwst-mid-infrared-instrument

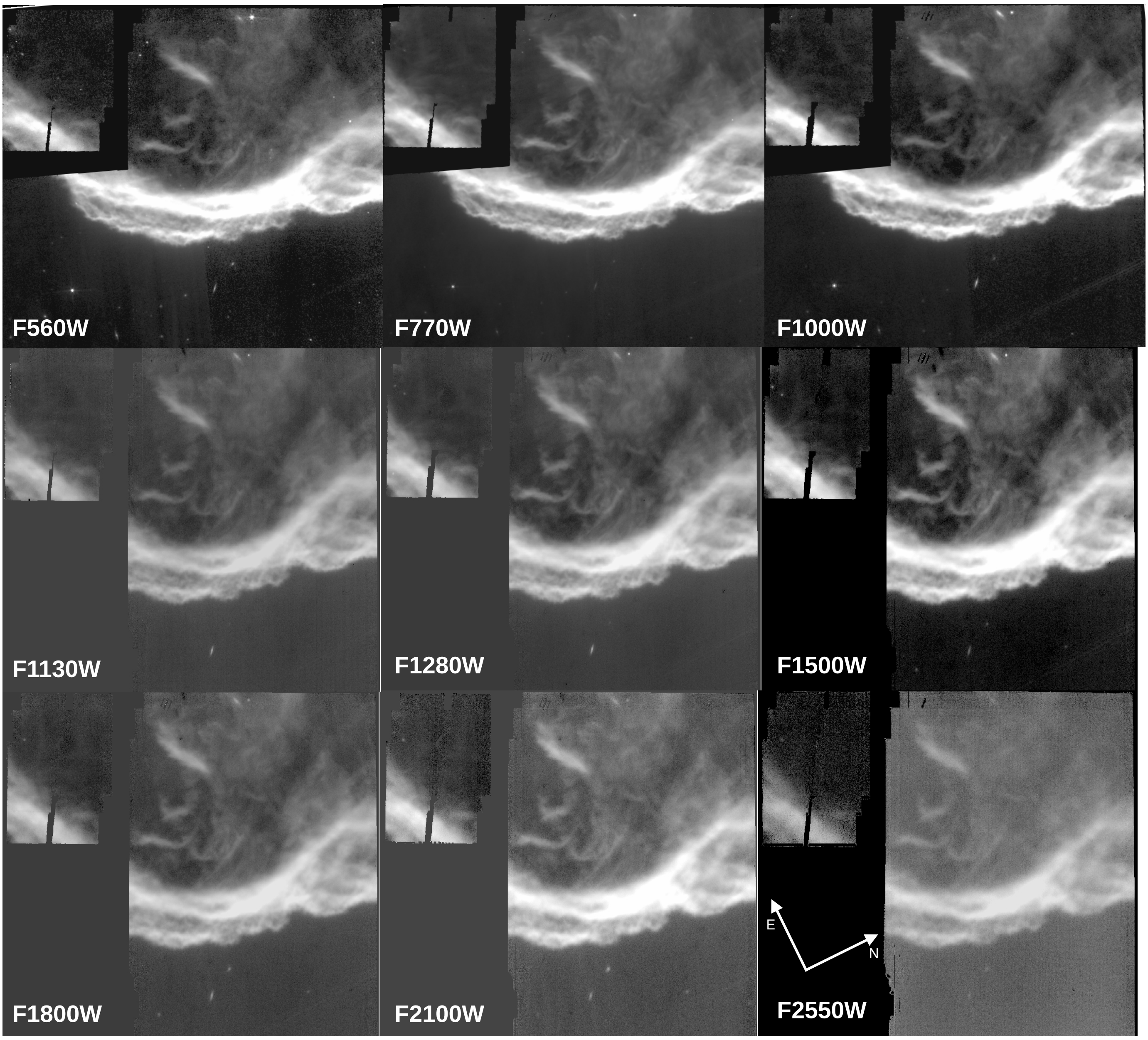

Fig. 4: PDR front in the nine MIRI filters. Left to right, from top to bottom: F560W, F770W, F1000W, F1130W, F1280W, F1500W, F1800W, F2100W, and F2550W. Logarithmic scales are used for all maps. Note: we have included the extra imaging coverage in the F560W, F770W, and F1000W filters provided by parallel imaging taken during MIRI IFU observations.

The MIRI instrument provides two readout modes, SLOWR1 and FASTR1, with frame times of 23.9 and 2.8 seconds, respectively. The recommendation was to use the FASTR1 read mode for MIRI observations. With the readout mode specified, combined with our ETC estimates, we specified 12 groups with a three-point CYCLING dither pattern, resulting in 100 seconds of total integration time per MIRI filter. We have also included the extra images taken in the F560W, F770W, and F1000W filters provided by parallel imaging during MIRI IFU observations. They have a higher S/N since the total integration time per filter is 866 s.

The MIRI data were processed entirely within the STScI pipeline environment. The coordinates of the Level 2 data were updated using GAIA DR3 prior to the Level 3 processing (mosaic) stage. The absolute flux calibration uncertainties are of the order of 3% for all filters (Gordon et al. in prep). We used the dedicated background observations except for the F560W and F770W filters since they are contaminated by faint stars and galaxies. For these two filters, the skymatch step was applied, which measures the sky statistics independently in the common region between images, and normalizes the relative sky differ-ences between all images to a reference image chosen from the list of input images. All MIRI images can be seen in Figs. 4.

### 3.3. HST

We are also using HST maps taken by the Wide Field Camera 3 (WFC3) IR camera in the broadband filters F110W and F160W, taken from the Hubble Heritage project[8]. These filters have bandwidths of 0.5 $\mu$m and 0.3 $\mu$m, respectively, mainly tracing the continuum due to the scattering of the incident radiation by large dust particles (e.g., Gordon et al. 2000; Draine 2003; Köhler et al. 2015). Their coordinates are also updated using the GAIA DR3 catalog. The F110W image is seen in Fig.5.

In addition, we utilize data from a companion HST program (PID 16747, PI Misselt) to estimate the H$\alpha$ emission using the Space Telescope Imaging Spectrograph (STIS). This program was designed to provide 1150-10000 Å spectra with the combined STIS slit apertures matched to the footprint of the JWST IFU observations. We extracted the profile of the H$\alpha$ flux across the PDR (shown in the left panel of Fig. 9) from the two-dimensional STIS G750L image by summing over the line in wavelength space.

### 3.4. Selected filters

Table 1 lists the main emission contributions in the 14 NIRCam, the 9 MIRI filters and the two HST/WFC3 filters we are using. The edge of the Horsehead is immersed in a large-scale H II region. The ionized gas is mainly traced by the F070W (H$\alpha$), F187N (Pa$\alpha$), and F405N filters (Br$\alpha$). These filters are also contaminated by the continuum, and the N II line and extended red emission (Witt & Lai 2020) in the F070W filter.

Molecular hydrogen ($H_2$) lines contribute to the F212N (1-0 S(1) at 2.12 $\mu$m), F323N (1-0 O(5) at 3.24 $\mu$m) and F470N (0-0 S(9) at 4.69 $\mu$m) filters; indeed those filters were designed to isolate the emission from the respective lines. The 1-0 S(1) line can be easily extracted from the F212N map by subtracting the continuum traced by the F210M filter. As shown by Habart et al. (2023) in the Orion Bar, two He I lines that are close to the 2.12 $\mu$m line are within the F212N spectral band, but their contribution is removed at the background correction stage (see §5.2). The 1-0 O(5) line is located on the shoulder of the 3.3 $\mu$m Aromatic Infrared Band (AIB), and the contribution of this band to the F323N filter is expected to be limited (5–15% across the observed area in the Orion Bar, see Habart et al. 2023), but not negligible. The 0-0 S(9) line is the only line expected to contribute to the F470N filter, however, a sufficiently accurate estimate of the continuum contribution is not possible using the two adjacent filters we have (F430M and F560W). Therefore, it is not possible to derive realistic images of the 1-0 O(5) and 0-0 S(9) lines without resorting to detailed modeling or incorporating an analysis of our IFU data here. This will be presented in a future paper. We note that all other broadband filters trace the scattering (below $\sim 2.5\,\mu$m) and the emission (above $\sim 2.5\,\mu$m) of grains in the PDR.

## 4. Photoevaporating flow from the illuminated edge

Figure 5 shows the two NIRCam and MIRI maps with the highest S/N, namely, in the F335M and F770W filters, together with NIRCam maps in filters sensitive to ionized hydrogen (i.e., F070W, F187N, and F405N) and the HST F110M map. Logarithmic scales are used and adjusted in order to see both the bright structures at the illuminated edge and the faint extended features in the IC 434 H II region. For all maps, the number of detected stars and background galaxies increases as the nebula becomes transparent, that is, in the H II region.

A network of faint striated features extending perpendicular to the PDR front into the H II region is also apparent in the F335M and F770W maps. The white arrows in the F770W map seen in Figure 5 show the typical length of these features ($\sim 25''$ or 0.05 pc). They are attributed to the photoevaporative flow of ionized gas from the neutral region to the H II region as the gas pressure in the H II region is not sufficient to contain the heated gas (Bertoldi 1989; Bertoldi & Draine 1996). The photoevaporating flow traced by the striated features is also visible in H$\alpha$ maps (e.g., Pound et al. 2003), in the NIRCam filters sensitive to ionized hydrogen (i.e., F070W, F187N, and F405N; see Fig. 5) and in the F110M HST map, though not as clearly as in the F335M and F770W bands because of lower S/Ns. It is actually not detected in the other broadband filters because of the noise. The observation of the photoevaporative flow in bands sensitive to nano-grains (F335M and F770W) and large grains (F110M) as well as in the ionized gas indicates that we may be detecting the entrainment of dust particles in the evaporative flow for the first time.

To make a more quantitative estimate of the potential nano-grain outflow, we first estimate possible contamination from higher $n$ ionized H lines in the F335M filter. There are 5 hydrogen recombination lines that fall within the F335M filter: Pfund ($n = 9$) and Humphreys ($n = 25, 24, 23, 22$). The NIRSpec spectrum from the H II region facing the Orion Bar (Fig. A.7 of Habart et al. 2023) shows that the contribution from other emission lines should be negligible. We take the measured H$\alpha$ line strength in the H II region from the HST STIS data (see Fig. 9) of $\sim 0.6 \times 10^{-17}$ erg s$^{-1}$ cm$^{-2}$ Å$^{-1}$ arcsec$^{-2}$, and Case B recombination ($T = 7500$ K, $n = 1000$ cm$^{-3}$, see Hummer & Storey 1987) to estimate the contribution of those 5 ionized H lines to the F335M signal which amounts to less than 1%. Therefore, no correction is applied for ionized line flux contamination. We estimate the source-free background in the F335M filter using regions well away from the front, bright sources, and the potential flow emission. The median pixel value in the flow is estimated by masking all pixels in the background subtracted image outside of the extended diffuse emission and inside the ionization front (see Fig. 5), resulting in a 'flow' region that extends 0.02-0.05 pc into the H II region with a median specific intensity of 0.08 MJy sr$^{-1}$. As we have no reliable constraint on the composition or size distribution of any dust component in the evaporative flow, we select plausible values for both. For the composition, we assume PAHs. We estimate the dominant PAH size that would contribute to the emission in the F335M filter by computing an emission weighted mean size. First, we compute the emission (e.g., Misselt et al. 2001) spectrum for a range of PAH sizes between 4-50 Å, typical of standard dust models, exposed to the $\sigma$-Orionis spectral energy distribution (derived from direct observations in HST program 16747) scaled to the PDR front (distance from $\sigma$-Orionis $\sim$3.5 pc). For each grain size, we integrate the emission across the F335M band response curve resulting in an estimate of the emission a single grain of that size would produce in the F335M bandpass. We define the emission weighted mean size as the size weighted by this size dependent emission function, resulting in an effective size of $\sim$8 Å. Note that in the $\sigma$-Orionis field at the position of the front, all grains smaller than $\sim$100 Å are stochastically heated. With the size and composition specifying the absorption efficiency (Li & Draine 2001), one can compute the emission (integrated over the stochastic temperature distribution of the grain). Assuming a density of 2.24 g cm$^{-3}$ and the median pixel flux density computed above, we arrive at a small dust grain mass of $\sim 1{\times}10^{-12} M_\odot$ per pixel or $\sim 5{\times}10^{-6} M_\odot$ in the flow. We note that the exact mass of dust in the flow is dependent on the size of the grain/PAH assumed; detailed modeling of the dust composition and size distribution is necessary to constrain the mass strongly. Pound et al. (2003) estimated the mass of the Horsehead to be $M(H_2){\sim}27\ M_\odot$ with a lifetime of $\sim 5\times10^6$ years assuming that the velocity of the evaporating flow is equal to the thermal sound speed in the ionized gas (10 ). If we assume the dust evaporative flow observed here is coupled to the gas flow, that corresponds to the flow propagating $\sim 10^{-5}$ pc year$^{-1}$ from the front. For the observed size of the flow (0.02-0.05 pc), this corresponds to a time-scale of 2-5$\times10^3$ years. This time-scale corresponds to a dust mass loss rate $\sim 1-3\times10^{-9} M_\odot$ year$^{-1}$. Assuming a typical gas-to-dust mass ratio of several $\times10^2$ in molecular clouds (Frisch & Slavin 2003; Compiègne et al. 2008), from the observed dust evaporative flow we derive a survival time for the Horsehead of $1-3\times10^6$ years. This is consistent with the Pound et al. (2003) estimate given the uncertainties in the calculated dust mass.

[8] NASA/ESA, and the Hubble Heritage Team (STScI /AURA)

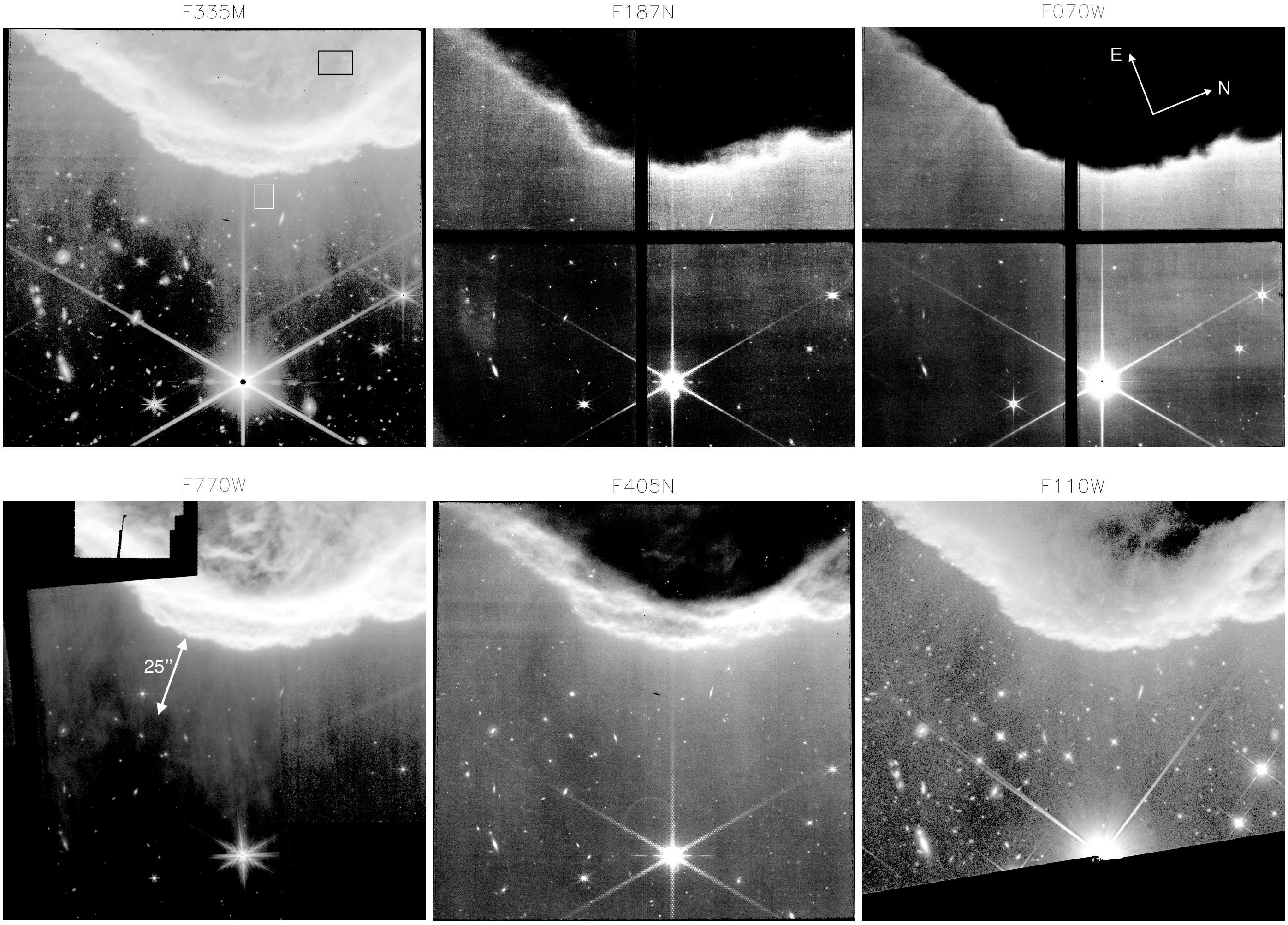

Fig. 5: NIRCam, MIRI and HST maps in the filters where a faint extended emission in the H II region IC 434 is detected. Logarithmic scales are used in order to display both the bright illuminated filaments in the PDR and the faint extended emission in the H II region. The MIRI F770W and HST F110W maps are projected on the NIRCam grid. Image width: 132″ or 0.26 pc. The NIRCam F187N and F070W images illustrate the pattern of the four short-wavelength detectors in constrast the single detector used at long wavelengths that covers the same region. The MIRI F770W image illustrates the coverage of the MIRI imager, including the "extra" small, neighboring field due to the Lyot coronagraphic field-of-view. The black rectangle in the F335M map shows the region where the ”background” brightness for the F070W, F090W, F187N, F405, and F110W filters is computed, and the white rectangle the region where it is computed for all other filters (see Sect. 5.2). The white line in the F770W image shows the typical length (∼ 25″ or 0.05 pc) of the detected striated features attributed to the photoevaporative flow from the illuminated edge (see Sect. 4). The north and east directions are seen on the F070W image.

## 5. Illuminated edge

Figures 2 and 3 present the NIRCam maps of the illuminated edge of the Horsehead in the bands where spectral lines are expected to contribute significantly, and in the bands dominated by continuum and IR features, respectively. Figure 4 presents the nine MIRI maps for the same region. For this first analysis, we have projected all maps (including the HST maps in the F110W and F160W filters) on the grid of the NIRCam LW maps using a bilinear interpolation, keeping all of them at their native angular resolution.

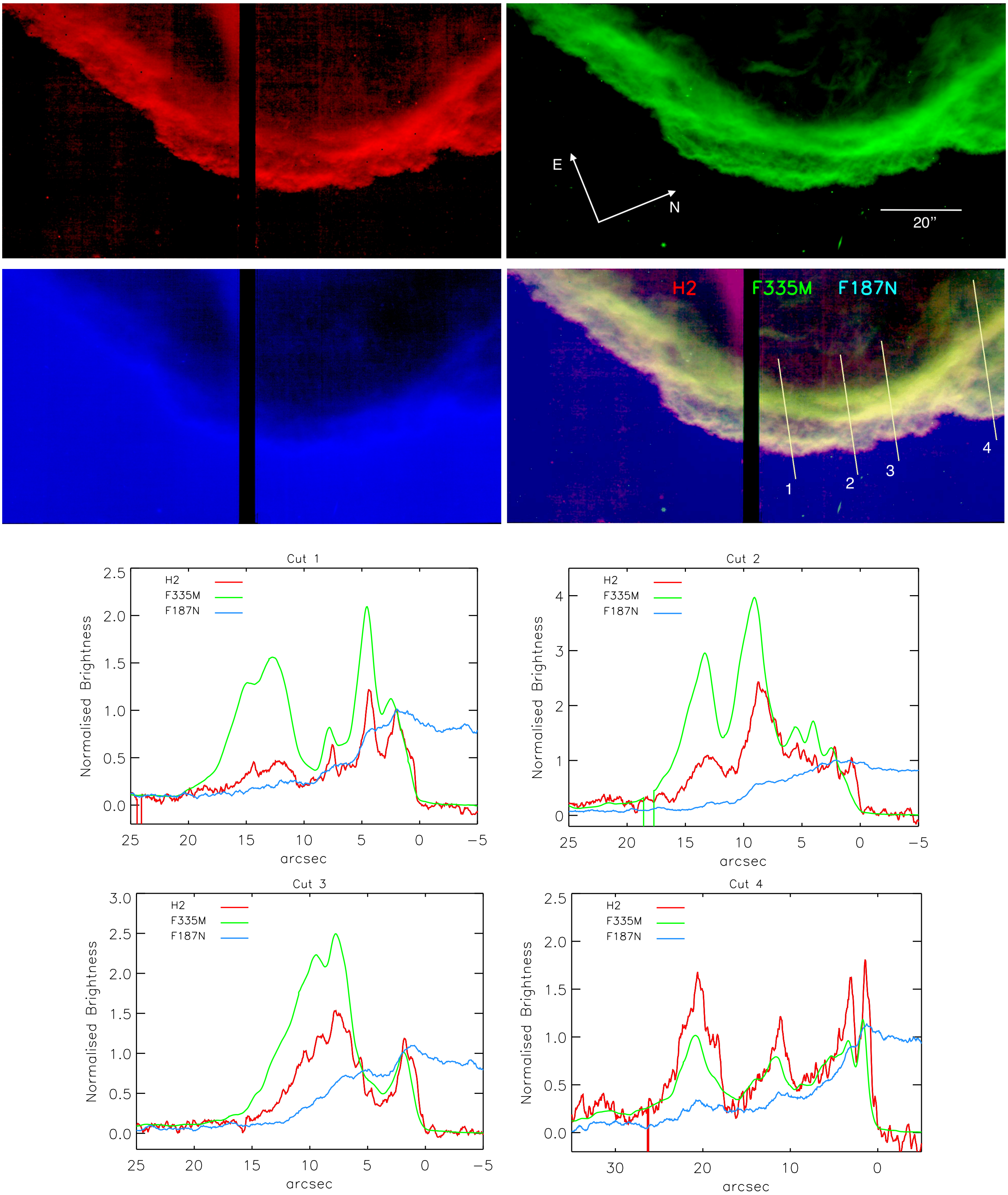

Fig. 6: Illuminated edge of the Horsehead in the $H_2$ line at 2.12 $\mu$m derived from the F212N and F210M filters (red), in the F335M (green) filter, and F187N (red) filter (top). Profiles of the relative brightness after background correction (normalized at a distance of 2″ from the edge) along the four cuts in the $H_2$ line at 2.12 $\mu$m (smoothed with a window size of 5 pixels, corresponding to $\sim$ 0.3″, in order to limit the noise), in the F335M filter and in the F187N filter (bottom). The three individual images are shown in logarithmic scales. They are used to make the RGB composite map. The four solid lines show the four cuts we use, labeled 1 to 4 from left to right. The position $z$ = 0″ in the profiles coincides with the more external edge of the PDR, as traced by the F335M filter (see §5.1). Negative values of $z$ are towards the illuminating star.

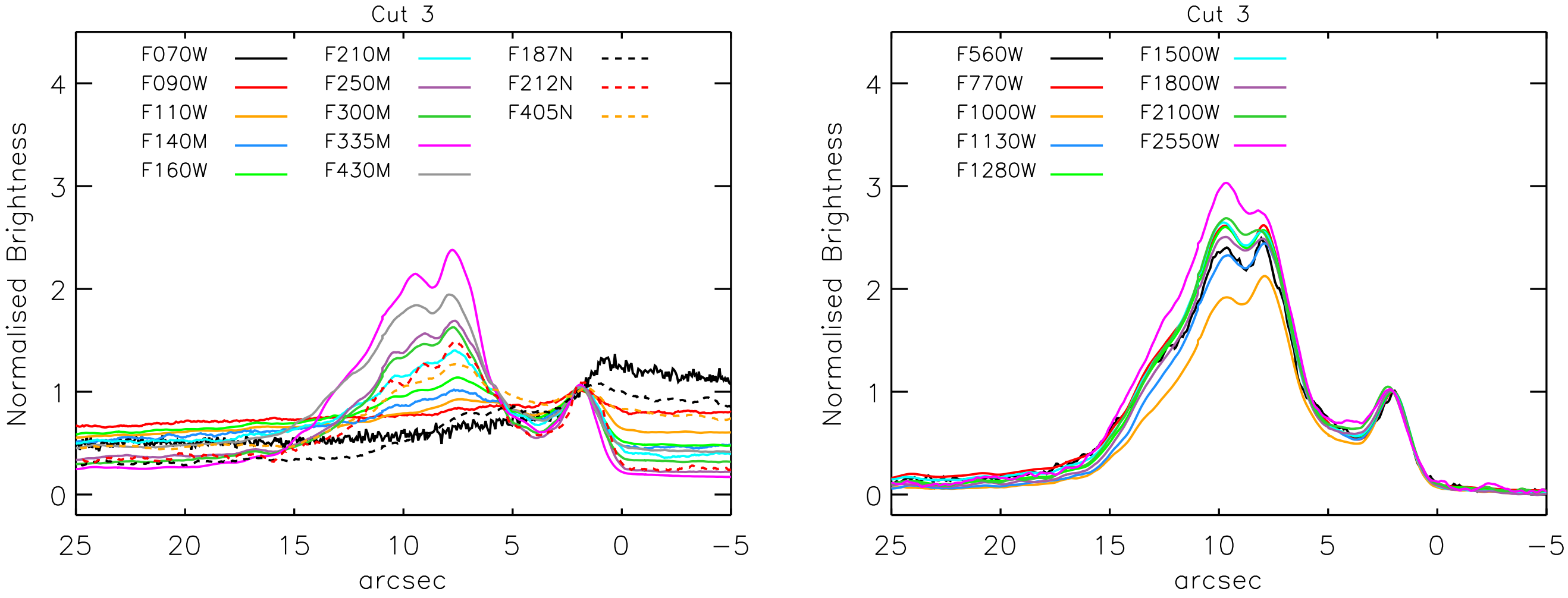

Fig. 7: Profiles of the relative brightness (normalized at a distance of 2″from the edge) for NIRCam and MIRI filters for cut 3 with no background correction.

### 5.1. External edge of the PDR

In the following sections, the images shown in Figs. 2 to 4 are analyzed together with brightness profiles across four selected cuts along the $\sigma$-Orionis direction and crossing the illuminated interface at different positions (as shown in Fig. 6). For all profiles, the $z$-axis is in the direction of $\sigma$-Orionis. The position $z = 0''$ coincides with the more external "edge" of the PDR which is empirically defined using the F335M map which has the highest S/N. It corresponds to the first position for each profile in the PDR where the emission is greater than 3 times the standard deviation in the H II region facing the PDR. The cut labeled "2" crosses two bright filaments at positions $z = 9''$ and $13''$ which appear to join at the position $z \sim 14''$ of the cut labeled "1". The cut labeled "3" crosses the center of the IFU footprint and at $z \sim 9''$ the peak position of the PDR in the F335M filter. The cut labeled "4" crosses a more complex region with three bright filaments.

### 5.2. Background correction

As illustrated in the brightness profiles shown in Fig. 7 for the third cut, the H II region in front of the illuminated edge, as well as the molecular region behind the illuminated edge, appear flat on large scales in all filters (note that for all images presented in Figs. 2 to 4 logarithmic scales are used and adjusted in order to evidence faint small-scale structures in the molecular regions). The averaged brightness in both regions is not zero because of possible contributions due to ionized gas in the F070W, F187N, and F405N filters, and dust emission or scattering in the other filters. All filters can also be affected by residual instrumental contamination. As a consequence, the bright filamentary structures detected in all filters at the illuminated edge appear on top of a flat "background", seen in the brightness profiles shown in Fig. 7. Therefore for the quantitative analysis of the brightness of these filamentary structures we need to subtract for all filters a "background" brightness.

For the filters tracing the ionized gas (F070W, F187N, and F405N) the background brightness is computed as the mean value inside a $10'' \times 7''$ rectangular window taken in the dense molecular regions where the brightness is minimal (shown as a black rectangle in the F335M map presented in Fig. 5). For the F090W and the F110W filters, the background brightness is computed in the same window since it appears that the brightness in the H II region is higher than in the dense region. For all other filters, it is computed as the mean brightness inside a $6'' \times 7''$ rectangular window in the H II region in front of the illuminated edge (shown as a white rectangle in the F335M map presented in Fig. 5). The windows used to calculate the background brightness are arbitrary, but we have verified that their precise position and size do not affect our results.

### 5.3. Spatial complexity of the edge of the Horsehead in the NIR and MIR

The figures 2 to 4 show multiple small-scale structures along the PDR and inside the dense cloud which are revealed in incredible detail in all filters. The top panel of Fig. 6 presents an RGB composite image with the maps in the F187N and F335M filters, and in the $H_2$ 1-0 S(1) line at 2.12 $\mu$m (computed with the difference between the F212N and F210M maps to subtract the spectral background and isolate the line emission). Four solid lines show the cuts we use to extract brightness profiles across the illuminated interface at the four selected positions. The bottom panels of Fig. 6 shows brightness profiles along these four cuts for the F187N and F335M filters, and the $H_2$ line. The F187N map mainly captures the distribution of ionized gas via the Pa$\alpha$ line, the F335M traces mostly the emitters of the 3.3-3.4 $\mu$m aromatic and aliphatic features and some continuum, and the $H_2$ map the distribution of warm $H_2$. The 3.3-3.4 $\mu$m features, and the $H_2$ line are sensitive to both the FUV radiation field and the gas density, so the interlaced bright and narrow filaments visible in the F335M and $H_2$ maps reveal the illuminated edge of dense structures.

Fig. 8 shows brightness profiles along the four cuts for all NIRCam filters tracing the ionized gas and the dust scattering and emission, and for MIRI filters which trace the dust emission. For all broadband filters tracing the dust emission (above $\sim 2.5\,\mu$m) the profiles appear remarkably self-similar, while the angular resolution (FWHM of the PSF) varies from $\sim$ 0.1 to $\sim$ 0.8″. In practice, for all filters, we do not detect any bright structure with a width of less than 1.5″, which corresponds to twice the angular resolution for the F2550M band. We conclude that most of the complexity of the spatial distribution in the NIR and MIR appears to be correctly resolved. There is one notable exception visible in the fourth cut due to a bright filament located just behind the edge ($z = 2 - 0''$) and which has a width around 1", so its amplitude decreases with increasing wavelengths above $\sim 12\,\mu$m due the decrease of angular resolution. If we exclude this filament, we can consider that for this first analysis, it is not necessary to convolve all the maps to the same angular resolution before performing a multi-wavelength analysis.

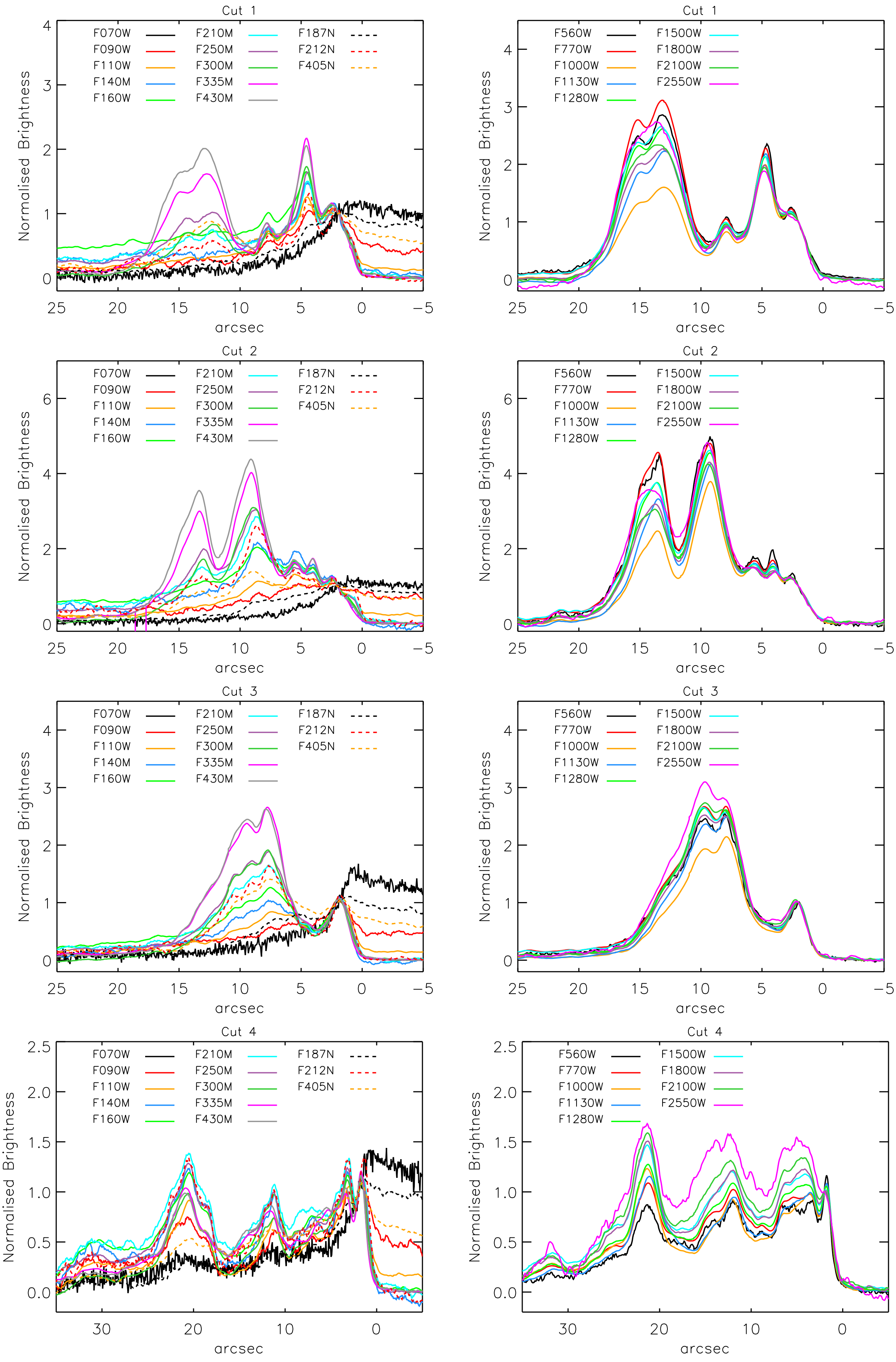

Fig. 8: Profiles of the relative brightness (normalized at a distance of 2″from the edge) for all filters after background correction (see §5.2).

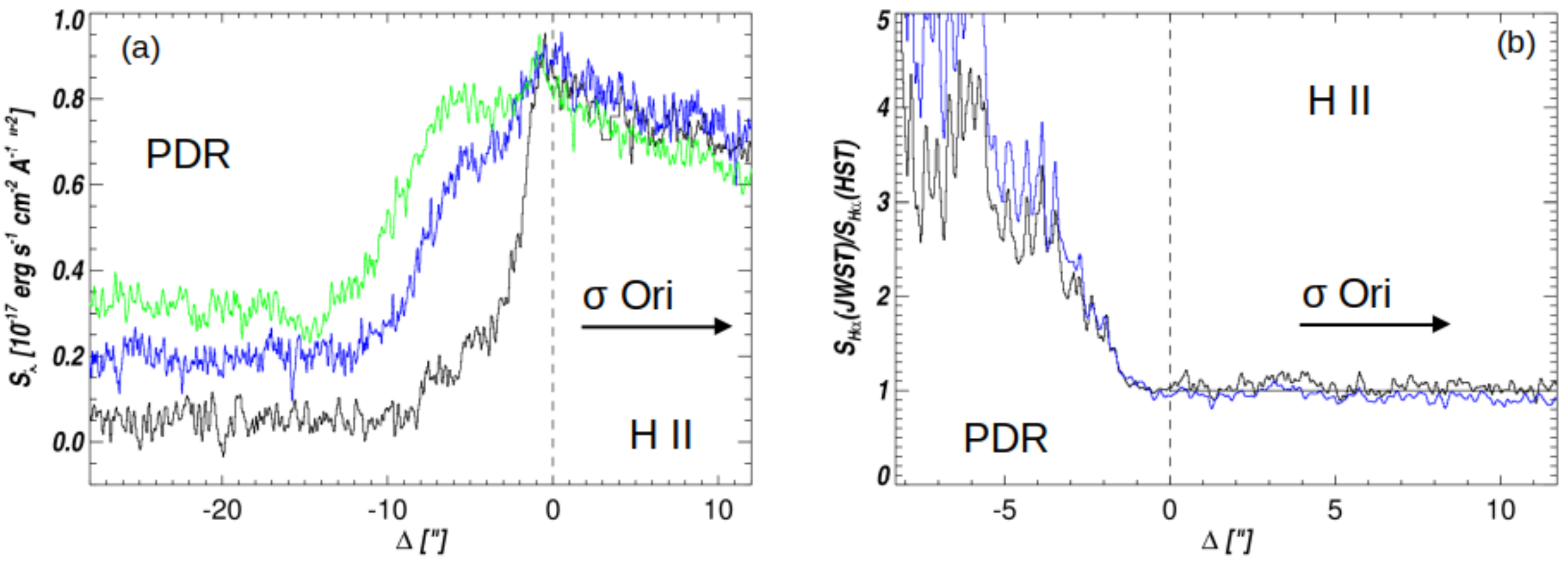

Fig. 9: Comparison of measured H$\alpha$ using HST to JWST estimates. (a) Profile of H $\alpha$ from HST STIS (black) across the front compared to that derived - see text - from F187N (blue) and F405N (green). (b) the ratio of H$\alpha$ emission estimated from JWST F187N (black) and F405N (blue) extracted from apertures matched exactly to HST STIS profile. In both plots, 0 on the x-axis corresponds to the edge of the PDR - see §5.1 - for profile cut 3 (which corresponds to the location of the JWST IFU and HST STIS long slit footprints).

## 6. Physical structure of the PDR

PDRs present a chemical layering at their illuminated surface (e.g., Hollenbach & Tielens 1999). The extreme-UV (EUV, $h\nu >$ 13.6 eV) photons from the illuminating massive stars produce a H II region facing the PDRs where the gas is fully ionized. The boundary between the H II region and the PDRs is delimited by the ionization front (IF), corresponding to the transition from ionized to neutral atomic hydrogen (H+/H). The position of the IF is generally taken at the peak of the H$\alpha$ emission. Deeper in the PDR, FUV ($6 < h\nu < 13.6$ eV) photons interact with neutral gas in the outer layers of the molecular clouds, photo-dissociating molecules. The dissociation front (DF) corresponds to the layer where the gas is half molecular and half atomic. It is traced here by the map of the $H_2$ 1-0 S(1) line at $2.12\,\mu$m.

### 6.1. Ionized gas

Our JWST images include three tracers of the IF, Pa$\alpha$ (F187N), Br$\alpha$ (F405N), and H$\alpha$ (F070W). In addition, we have a direct measure of the H$\alpha$ emission over a restricted region of the Horsehead PDR from a companion HST STIS program (PID 16747). In that program, the STIS slit was placed to overlap exactly with the JWST NIRSpec and MIRI IFU mosaics (see Fig. 1). Practically, the F070W image will contain significant contamination from N II, ERE, and the scattered light continuum owing to its wide bandpass; thus, we focus on the narrowband to trace the IF and check the consistency of the tracers. To estimate the Pa$\alpha$ and Br$\alpha$ line flux images from the F187N and F405N filters, we first corrected the images for a uniform foreground extinction along the line of sight to the Horsehead with an E(B-V) = 0.05 (Lee 1968) and a standard $R_V$=3.1 extinction curve (Fitzpatrick & Massa 2009). We then performed a background subtraction using a linear interpolation between adjacent medium bands (F140M/F210M for F187N and F250M/F430M for F405N).

To identify the position of the IF, we can scale the Pa$\alpha$ (F187N) and Br$\alpha$ (F405N) images to predict the H$\alpha$ emission using case B recombination and compare to the directly observed H$\alpha$ emission from HST-STIS. The exact parameters of the case B recombination scaling used are not critical and here we assume scaling appropriate for $n$ = 1000 cm$^{-3}$, $T$ =7500 K: 0.124 for Pa$\alpha$ and 0.030 for Br$\alpha$ (Hummer & Storey 1987). We scale the F187N and F405N by those factors and extract sub-images matched exactly to the STIS slit, then extract a profile cut as shown in Fig. 9. While the observed H$\alpha$ emission and that inferred from case B and our JWST data agree very well in the H II region, they deviate significantly as we cross the ionization front into the neutral region with the longer wavelength Br$\alpha$ prediction being higher than the Pa$\alpha$ prediction which is, in turn, higher than the observed H$\alpha$ emission. This observation is consistent with an optical depth effect due to the longer wavelengths being less impacted by dust extinction. If the observed H$\alpha$, F187N, and F405N emissions are not only direct emission from the front but includes a fraction that must pass through the front itself, the longer wavelengths will be less attenuated and, thus, they appear brighter and imply a higher H$\alpha$ flux.

Such a differential attenuation could indicate that the exciting star illuminates the front at an oblique angle, bringing significant illumination from behind the PDR as illustrated in Fig. 13. Note that the inclination of $\sigma$-Orionis has been suggested by Habart et al. (2005) in order to explain $H_2$, $^{12}$CO, C$^{18}$O and dust (aromatic emission at 7.7 $\mu$m and 1.2 mm continuum) observations of the illuminated edge of the Horsehead, with an angle estimated to be $\sim 6^\circ$ from the plane of the sky. The Horsehead nebula is seen in silhouette against the background and much of the background emission is from the illuminated surface of the PDR behind the Horsehead. We show in §8 that attenuation effects strongly impact the brightness observed in all filters tracing the dust emission or scattering.

The ionized gas is traced by the F070W, F187N and F405N filters. The different estimators of excited H emission - H$\alpha$, Pa$\alpha$, and Br$\alpha$ - and their sharp transition at the molecular cloud allow us to localize the position of the IF within a fraction of $1''$ (approximated by the vertical dashed line in Fig. 9). An interesting result is that the continuum subtraction we have performed on the F187N and F405N maps does not change the peak positions. We can therefore consider that the peak positions for these two maps, without any correction, nicely give us the location of the IF.

### 6.2. Molecular gas

The map of the ro-vibrational line emission at 2.12 $\mu$m (see Fig. 6) is derived from the F212N and F210M maps (see §3.4). It traces the molecular hydrogen excited by the FUV radiation field received at the edge of molecular structures. At the edge of a molecular cloud illuminated edge-on, the $H_2$ emission first increases mainly because of the increase of the density, then falls inside the molecular cloud because of the decrease of the FUV radiation field due to both $H_2$ self-shielding and dust extinction of the radiation of the illuminating star. This explains the filamentary structures of the spatial distribution of the $H_2$ emission visible along the illuminated edge. Each filament traces the illuminated edge of a dense molecular structure.

As can be seen in Fig. 6, the spatial distribution of the $H_2$ line is in fact more complex, displaying numerous sub-structures on scales smaller than that of the brighter filaments. The smallest ones have sizes around 1", and they were not resolved with the previous mapping of the 2.12 $\mu$m $H_2$ line by Habart et al. (2005). Here, we are not observing the illuminated edge of one simple molecular structure, but the superposition of the illuminated edges of several molecular condensations which are not observed at the same position.

The maps of the nano-grain emission present a spatial structure comparable to the $H_2$ map, as illustrated in Fig. 6 for the F335M map. This result is fully expected, as the nano-grain emission depends on the gas density and the local radiation field in a way that is comparable to the $H_2$ emission. Both the $H_2$ and nano-grain emissions come for the illuminated edges of dense structures. However, the F335M and $H_2$ maps are not fully identical, and there are at least two interesting differences.

Firstly, just behind the external edge of the PDR (defined in Sect. 5.1), namely, at a distance of $z \leq 1-2''$ ($1''$ corresponding to $2\times10^{-3}$ pc or 400 au), the brightness profiles for the four cuts (bottom panels of Fig. 6) show that the increase of the $H_2$ emission is significantly sharper than for the F335M filter. This effect is even more pronounced for cuts 2 and 3. The same result is found when the $H_2$ map is compared to all maps in filters tracing the nano-grain emission or the dust scattering: just behind the edge, it is always in the $H_2$ filter that the illuminated edge appears the sharpest. This results in an $H_2$ emission excess just behind the illuminated edge, as illustrated in the composite maps of Fig. 6 (with the $H_2$, F335M, and F187N maps in red, green, and blue, repectively) and Fig.10 (with the $H_2$ and F210M maps in red and blue, respectively). Note that the $H_2$ excess appears all along the external edge of the PDR, where numerous small scale features which look like Rayleigh-Taylor instabilities are detected. Detailed modeling is obviously necessary to understand the physical processes at the origin of this excess ($H_2$ formation, excitation and dissociation, variation of dust properties, etc).

Secondly, for for the first three cuts and at larger distances ($z > 2''$), the F335M brightness increases more inside the PDR than the $H_2$ brightness (note that all profiles in Fig. 6 are normalized at a distance of $2''$from the edge). This results in strong color variations in the composite map shown in Fig. 6: green filaments with low values of the $H_2$/F335M ratio are visible in the internal parts, while red structures with higher $H_2$/F335M ratio are seen behind the edge. Such color variations are likely due to dust attenuation which impacts the $H_2$ line at 2.12 $\mu$m more than the nano-grain emission around 3.35 $\mu$m. The total column density inside the PDR increases with increasing distance from the edge (e.g., Schirmer et al. 2020; Hernández-Vera et al. 2023), thus the column density of the material responsible for the attenuation increases with increasing distance. Kaplan et al. (2021) have derived, using NIR observations of different excited rovibrational $H_2$ levels from the same upper $v$ and J state and around the peak position of our third cut ($z \sim 10''$), a value of the extinction magnitudes $A_K = 0.7$ in the K-band (2.2 $\mu$m), corresponding to $A_V \sim 7$. We show in §7 that the dust attenuation has also a strong impact on the observed nano-dust emission and large grain scattering in all filters.

### 6.3. Ionization and dissociation fronts

We see on the four brightness profiles shown in Fig. 6 that the first peak of the $H_2$ emission (tracing the DF location), located just behind the external edge of the PDR, spatially coincides with the F187N peak (tracing the IF location, see §6.1). The upper limit of the distance between the two peaks is $\sim 0.25''$. We can conclude (1) that along the edge of the Horsehead, the IF and the DF appear to be located at distances $\sim 1-2''$ from the external edge seen in the images shown in Fig. 6, and (2) that the distance between the two fronts $d_{IF-DF}$, corresponding to the thickness of the atomic layer, is below $0.25''$, corresponding to $5\times10^{-4}$ pc = 100 au. From the H$\alpha$ map taken by Pound et al. (2003) with the 0.9 m telescope at the Kitt Peak National Observatory, and the map of the 2.12 $\mu$m $H_2$ line taken by Habart et al. (2005) at the 3.6 m New Technology Telescope, Hernández-Vera et al. (2023) have recently inferred an upper limit for $d_{IF-DF}$ of 650 au. We find a lower limit because of the higher angular resolution of the JWST, especially for the $H_2$ line at 2.12 $\mu$m which allows us to resolve the $H_2$ filaments.

Hernández-Vera et al. (2023) have also shown that a thickness of the atomic layer of a few hundred au can be reproduced by isobaric stationary models with relatively high thermal gas pressures ($2-4\times10^6$ K cm$^{-3}$) consistent with estimates deduced from the physical conditions of the gas and the excitation. However, they also observed that the thermal gas pressure is higher in the PDR interface than in the H II region, suggesting the gas is compressed; this is compatible with the detection of photo-evaporated matter in the H II region (see §4).

The advection of the gas across the ionization front due to the photo-evaporation brings the H/$H_2$ transition closer to the ionisation front than in the static case (e.g., Bertoldi & Draine 1996). The effects on the location of the H/$H_2$ transition have been analyzed by Maillard et al. (2021) using a semi-analytical model in a 1D plane-parallel PDR. For high-excitation PDRs such as the

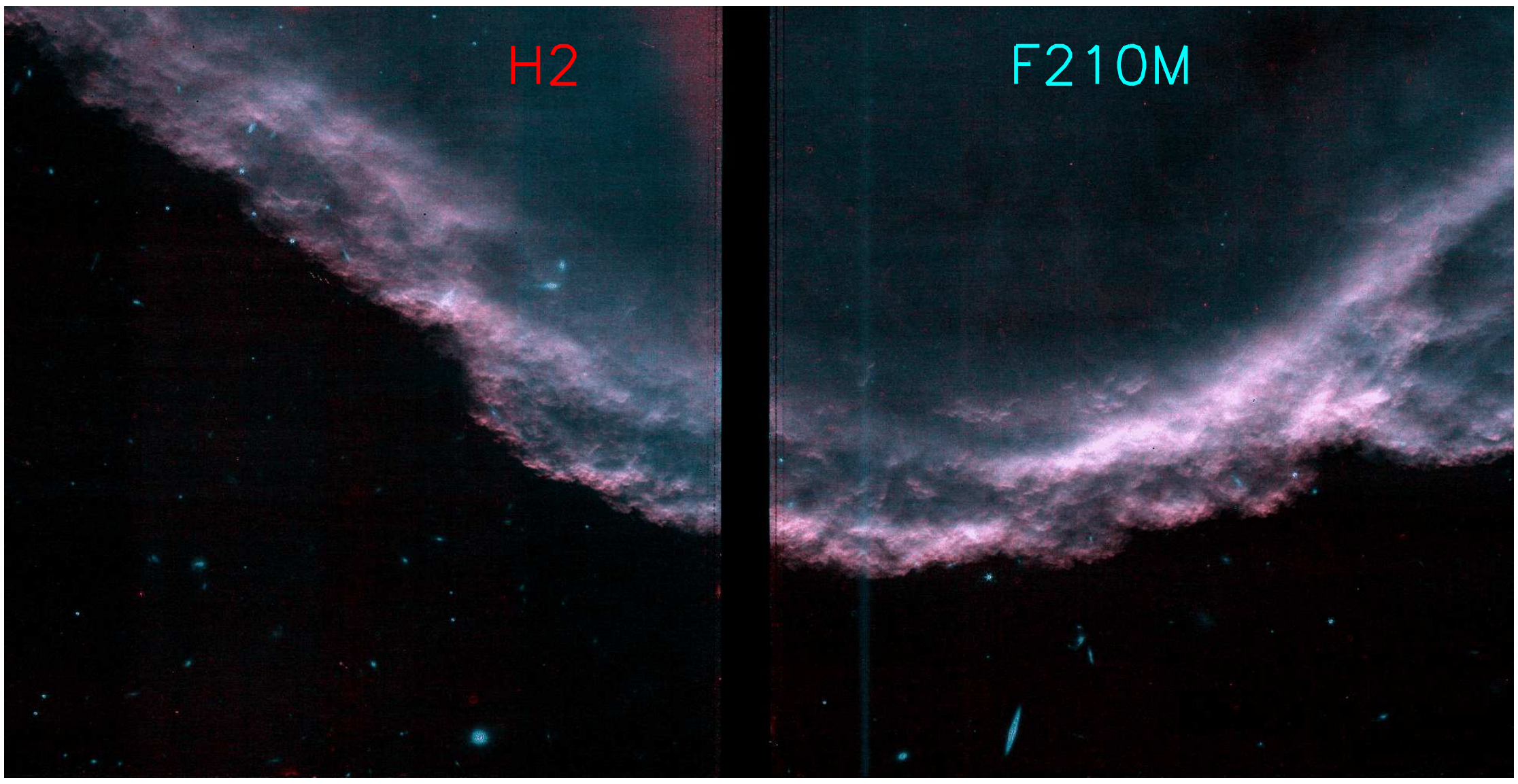

Fig. 10: Illuminated edge of the Horsehead in the 2.12 $\mu$m $H_2$ line (red) and the F210M filter (blue).

Orion Bar, the thickness of the atomic layer does not appear to be stronly reduced. But for low-excitation PDRs illuminated by O-stars such as the Horsehead, the dissociation and ionization fronts can merge and the atomic neutral region fully disappears. Our observations are compatible with such a scenario. However, it is also necessary to take into account the in-situ formation of $H_2$ by various possible mechanisms. As discussed by Wakelam et al. (2017), the classical picture of $H_2$ formation is the physisorption of a H atom on a grain surface which then reacts with a second from the gas phase (Eley-Rideal mechanism) or with another on the grain surface (Langmuir-Hinshelwood mechanism) before reacting to form an $H_2$ molecule. Bron et al. (2014) have shown that at the illuminated edges of PDRs the Langmuir-Hinshelwood mechanism on stochastically-heated nano-grains is as efficient as the Eley-Rideal mechanism. Jones & Habart (2015) have proposed an alternative mechanism to form $H_2$ in PDRs with low to moderate radiation fields ($1 \leq G_0 \leq 100$) via the photolysis of hydrogenated amorphous carbon nano-grains. In any case, a detailed modeling is necessary to give any definitive conclusion.

## 7. Dust emission and scattering

The dust emission is traced by all broadband filters above $\sim 2\,\mu$m, while broadband filters in the spectral range $1 - 2.5\,\mu$m are dominated by the scattering of the incident radiation by large grains. The images in all broadband NIRCam and MIRI filters look quite similar (see Figs. 3 and 4). However the brightness profiles (as shown for the third cut in the top panels of Fig. 12) present significant differences between filters, also evidenced in the color ratio profiles (middle panels of Fig. 12). Color ratios are calculated relative to the F770W map, which corresponds to the spectral band where extinction should be minimal. Hereafter, we mainly focus on the third cut in the analysis. Plots for the other cuts are presented in the Appendix (Figs. A.1 and A.2).

Almost all color ratios decrease as the distance from the edge increases beyond 5″. For NIRCam filters, the amplitude of this decrease is larger for shorter wavelengths: the F070W/F770W ratio presents the highest decrease, and the F430M/F770W the smallest one. For the MIRI filters the strongest decrease inside the PDR is found for the F1000W/F770W ratio, and the amplitude gets weaker for F1130W/F770W and F1800W/F770W (in that order). The amplitude is still lower for the F1280W/F770W, the F2100W/F770W, and the F1500W/F770W ratios. The F560W/F770W and F2550W/F770W ratios appear almost flat, with some fluctuations. Comparable results are obtained through the first two cuts (see Figs. A.1 and A.2 shown in the appendix), but not through the fourth cut. Spatial variations in color ratios translate into dramatic color variations in the RGB map of Fig. 11, with the F140M, F335M, and F770W maps in red, green and blue, respectively: blue/green filaments are visible inside the PDR, around 10″, due to a F140M deficit compared to the F335M and F770W, while red structures corresponding to a F140M excess are visible just behind the edge.

These spatial variations of the color ratio can simply be explained by dust attenuation, in the scenario discussed in §6.1 where the illuminating star is slightly inclined compared to the plane of the sky, bringing significant illumination from behind the PDR (as illustrated in Fig. 13). Indeed, the extinction curve longward of 1 $\mu$m exhibits a $\lambda^{-\alpha}$ behavior with a value of $\alpha = 1.6 - 1.8$, where $\alpha$ is a function of the ratio of total-to-selective extinction, $R_V$ (Fitzpatrick & Massa 2009). This can explain that the amplitude of the decrease of the color ratios gets systematically larger toward shorter wavelengths, down to 5 $\mu$m. Moreover, the extinction curve presents extinction excess around 9.7 $\mu$m and 18 $\mu$m due to Si–O stretching and O–Si–O bending modes, respectively (e.g., Weingartner & Draine 2001), which can explain why the stronger decrease is found for the F1000W/F770W ratio. Finally, the extinction curve also presents a flattening in the MIR spectral range (Lutz et al. 1996), which can explain why the F560W/F770W ratio appears almost flat.

The brightness profiles across the fourth cut do not present any comparable trend. This is likely due to different illuminating conditions. The first three cuts cross material illuminated from the back side of the nebula, so the observed emission is attenuated by material located in the front, in the direction of the JWST. A possible local geometry for these cuts could be a structure in the shape of a terraced field illuminated from the back side of the nebula, as illustrated in Fig. 13. At the contrary, the fourth cut does not appear to be dominated by attenuation effects. It should cross matter illuminated from the front side of the nebula, possibly also due to successive structures in the shape of a terraced field but seen from the front. But the real geometry can obviously be very complex.

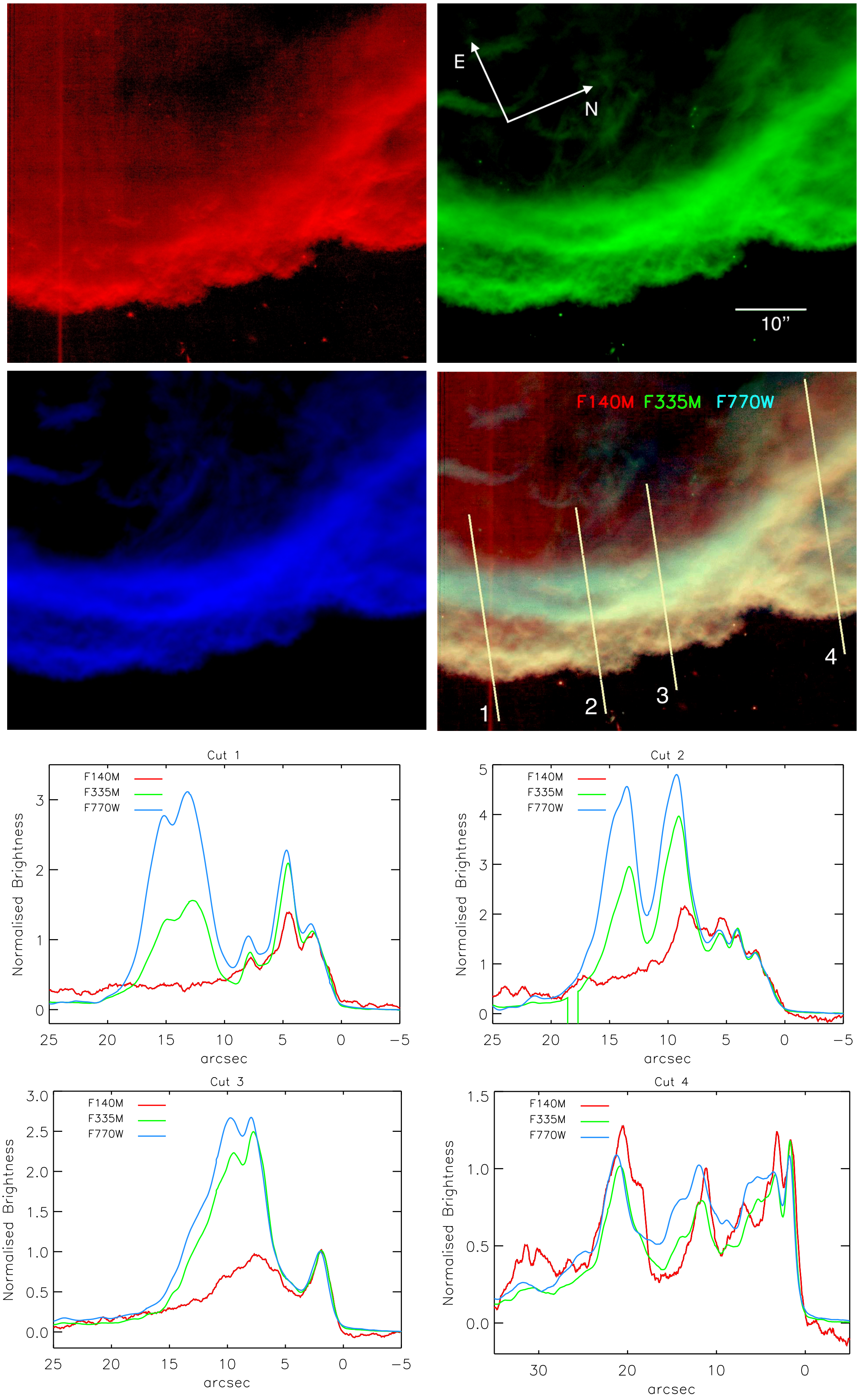

Fig. 11: Same as Fig. 6 for the F140M (red), F335M (green) and F770W (blue) filters.

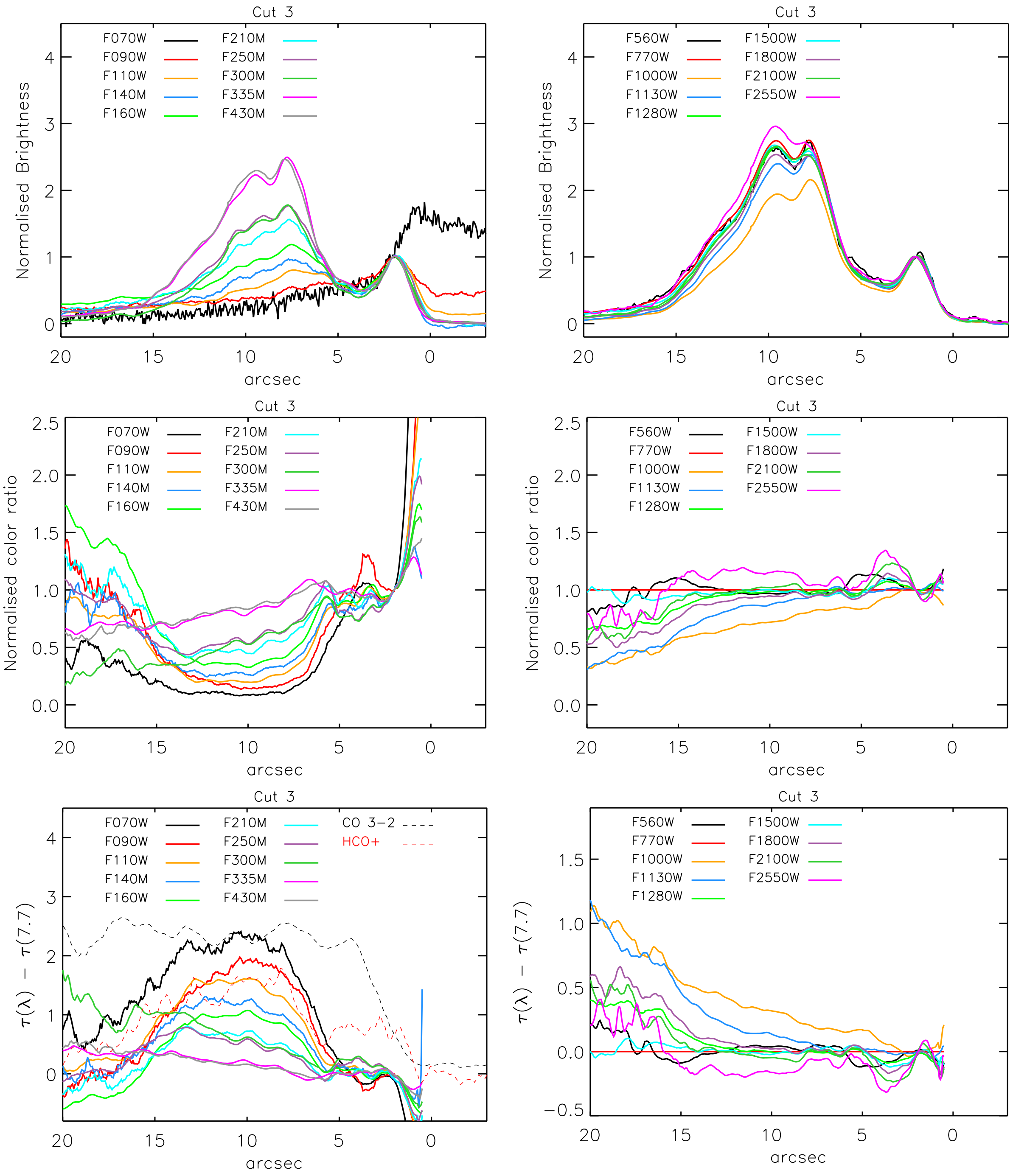

Fig. 12: Profiles of the relative brightness (normalized to 1 at a distance of 2″) for all broadband filters, along cut 3 (see Fig. 6) shown at the top. Middle panels: Normalized color ratios (normalized to 1 at a distance of 2″) relative to the F770W filter, smoothed with a window size of 10 pixels, corresponding to ∼ 0.6″, in order to limit the noise. Above ∼ 15″ the ratios are more noisy because the brightness is near 0. Bottom panels: Profiles of the optical depth difference along the third cut (see §8). On the bottom left panel is also plotted the normalized velocity-integrated intensity profiles of the CO $J = 3 - 2$ and $HCO^+$ $J = 4 - 3$ lines observed with ALMA by Hernández-Vera et al. (2023).

For filters below 2.8 $\mu$m, dominated by light scattered by large grains, the decrease of the color ratios with increasing distance (see the middle left panel of Fig. 12) stops around 12″ for cut 3 (around 15″ for cuts 1 and 2; see Figs. A.1 and A.2), then the color ratios increase with increasing distance. Such trend is not detectable for filters above 2.8 $\mu$m which are dominated by nano-grain emission. It could be due to an excess of scattered light resulting from multiple scattering by large grains located inside the PDR and illuminated by $\sigma$-Orionis. It could also be due to nano-grain depletion in the internal regions, from around 12″ from the external edge. Indeed, the modeling of Spitzer and Herschel observations by Schirmer et al. (2020) of the illuminated edge of the Horsehead have evidenced a nano-grain depletion relative to the gas by a factor 6-10 times compared to the diffuse ISM, but no spatial variation of this depletion was detected across the edge because of the limited angular resolution. With the JWST, we may have resolved this spatial variation. However, detailed modeling, including the anisotropy of the scattering which is forward peaked (e.g., Henyey & Greenstein 1941; Baes et al. 2022) and radiative transfer is needed to reach any definitive conclusion.

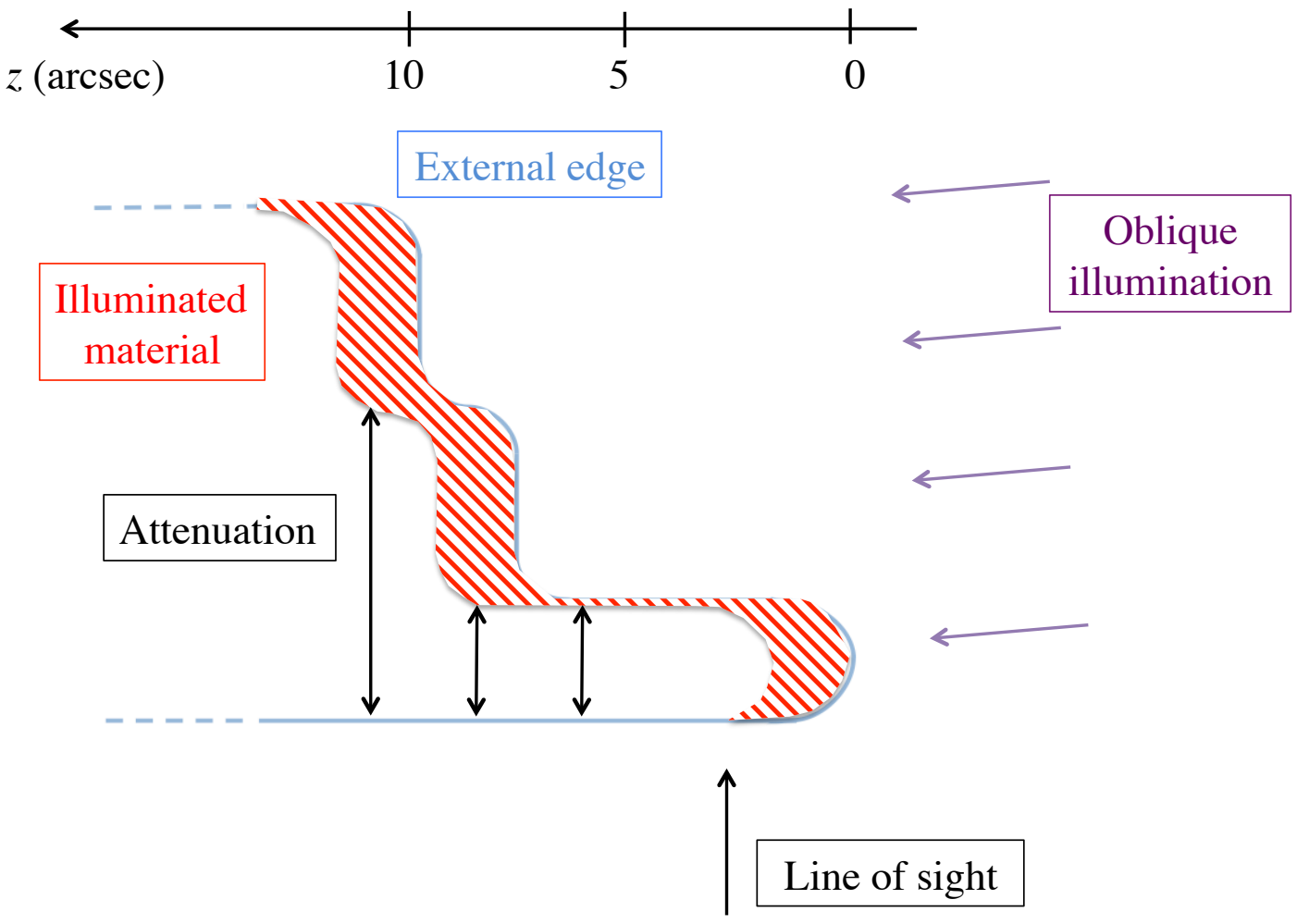

Fig. 13: Schematic view of the simple model used to quantify the attenuation in cuts 1 to 3.

## 8. Quantitative analysis of the dust attenuation

A detailed modeling of the observations is beyond the scope of this paper. However, we propose a first quantitative analysis of the attenuation using a very simple model to make a quantitative check to see whether dust attenuation could indeed be the dominant process explaining the large-scale color variations we observe here.

### *8.1. Simple model of the Horsehead*

In this model (see the schematic view in Fig. 13), we assume that the observed brightness is due to emission or scattering from a thin layer located behind the PDR illuminated with an oblique angle, which is attenuated by the material located between the illuminated layer and the observer. It is important to note that the scattering and emission we observe are due to material located in a thin layer ($A_V < 1$) just behind the illuminated edge (red regions in Fig. 13), while the dust responsible for the attenuation is located in the internal parts of the PDR (black arrows in Fig. 13). Any attenuation effect inside the illuminated layer is neglected. With this model, the observed brightness at wavelength $\lambda$ along a cut for the projected distance $z$ from the illuminated edge can be written as:

$$I(\lambda, z) = I_0(\lambda, z)\; e^{-\tau(\lambda,z)} \,, \tag{1}$$

where $I_0(\lambda, z)$ is the brightness that would have been observed without attenuation and $\tau(\lambda, z)$ the optical depth between the illuminated material and the observer along the line of sight.

We also assume that the spectrum of the emitted and scattered radiation is the same at all positions on the illuminated surface of the PDR, ignoring possible variations of the dust properties (abundance, size distribution, ionization state, ...), so that $I_0(\lambda, z)$ can be separated in two terms depending on $\lambda$ and $z$, respectively:

$$I_0(\lambda, z) = I_0(\lambda)\; \eta(z). \tag{2}$$

Here, $I_0(\lambda)$ is the brightness at wavelength $\lambda$ observed just behind the illuminated edge at a position for which we can neglect the attenuation, and $\eta(z)$ a multiplicative factor which depends on the column density of the emitting material illuminated for the projected distance $z$. Practically, for each cut, we take for $I_0(\lambda)$ the brightness measured at 2″ from the edge. Equations 1 and 2 allow us to compute the optical depth difference between wavelenghts $\lambda$ and $\lambda_{ref}$:

$$\begin{aligned}\Delta\tau(\lambda, z) &= \tau(\lambda, z) - \tau(\lambda_{ref}, z) \\ &= -\ln\left[\frac{I(\lambda, z)}{I_0(\lambda)} \times \frac{I_0(\lambda_{ref})}{I(\lambda_{ref}, z)}\right].\end{aligned} \tag{3}$$

Using a reference wavelength $\lambda_{ref}$ removes the unknown factor $\eta(z)$.

### *8.2. Quantitative analysis of the attenuation*

The bottom panels in Fig. 12 present the resulting profiles of optical depth differences $\Delta\tau(\lambda, z)$ computed for the third cut, with $\lambda_{ref} = 7.7\,\mu$m. Comparable results are obtained for the first and second cuts (see Figs. A.1 and A.2 in the appendix). The main sources of error on $\Delta\tau$ is our model and not the calibration uncertainties since $\Delta\tau$ is calculated from brightness ratios at given wavelengths (see Eq. 3). Consequently, in this first analysis, it is not possible to quantify the error on $\Delta\tau$. The $\Delta\tau$ profiles show many fluctuations due to our oversimplified model. For instance, for NIRCam filters below 2.8 $\mu$m, $\Delta\tau$ decreases in the third cut for distances higher than $\sim$ 12″, which may be due to variations of the emitted or scattered radiations discussed in §7. But here we focus on the general trend seen in Fig. 12, which is an increase of $\Delta\tau$ with increasing distance $z$ beyond 5″, with amplitudes depending on the wavelength, which is obviously related to the systematic decrease of the color ratios with increasing distance.

Figure 14 shows the spectral variations of $\Delta\tau(\lambda)$, computed for the third cut at the peak position $z_0$ in the F770W filter (see Fig. A.3 in the appendix the spectral variations of $\Delta\tau(\lambda)$ for cuts 1 and 2). They are compared to the spectral variations of $\Delta\tau_{ext}(\lambda)$, the synthetic optical depth difference computed for different extinction curves

$$\Delta\tau_{ext}(\lambda) = (\sigma_{ext}(\lambda) - \sigma_{ext}(\lambda_{ref})) \times N_H(z_0)\,. \tag{4}$$

Here, $\sigma(\lambda)$ is the extinction cross section per H atom at wavelength $\lambda$ and $N_H(z_0)$ the column density along the line of sight of the material responsible for the extinction for the projected distance $z_0$. For this comparison, we have adjusted $N_H(z_0)$ so that $\Delta\tau_{ext}(\lambda) = \Delta\tau(\lambda, z_0)$ at 10 $\mu$m. We note that $\Delta\tau(\lambda)$ measures the wavelength dependence of attenuation and not the extinction as such. Nonetheless, we find a relatively good match, but we also see several differences, along with the fact that no synthetic curve precisely matches the data on the full spectral range. Around the extinction features at 10 and 18 $\mu$m, the synthetic curve computed using the extinction law of Gordon et al. (2023) with $R_V = 3.1$ appears to be in good agreement with the data, but the curve with $R_V = 5.5$ cannot be ruled out at that stage. In the NIR spectral range, $\Delta\tau(\lambda)$ appears to increase at shorter wavelengths less steeply compared to the synthetic curves. This may reflect the increasing contribution due to scattering as the dust albedo increases towards shorter wavelengths (thus increasing the observed intensities at the shorter wavelengths), or the decreasing abundance of nano-grains. However, we must keep in mind that our model used to derive $\Delta\tau_{ext}$ is very simple (likely too simple) and our goal here is not to obtain a perfect match.

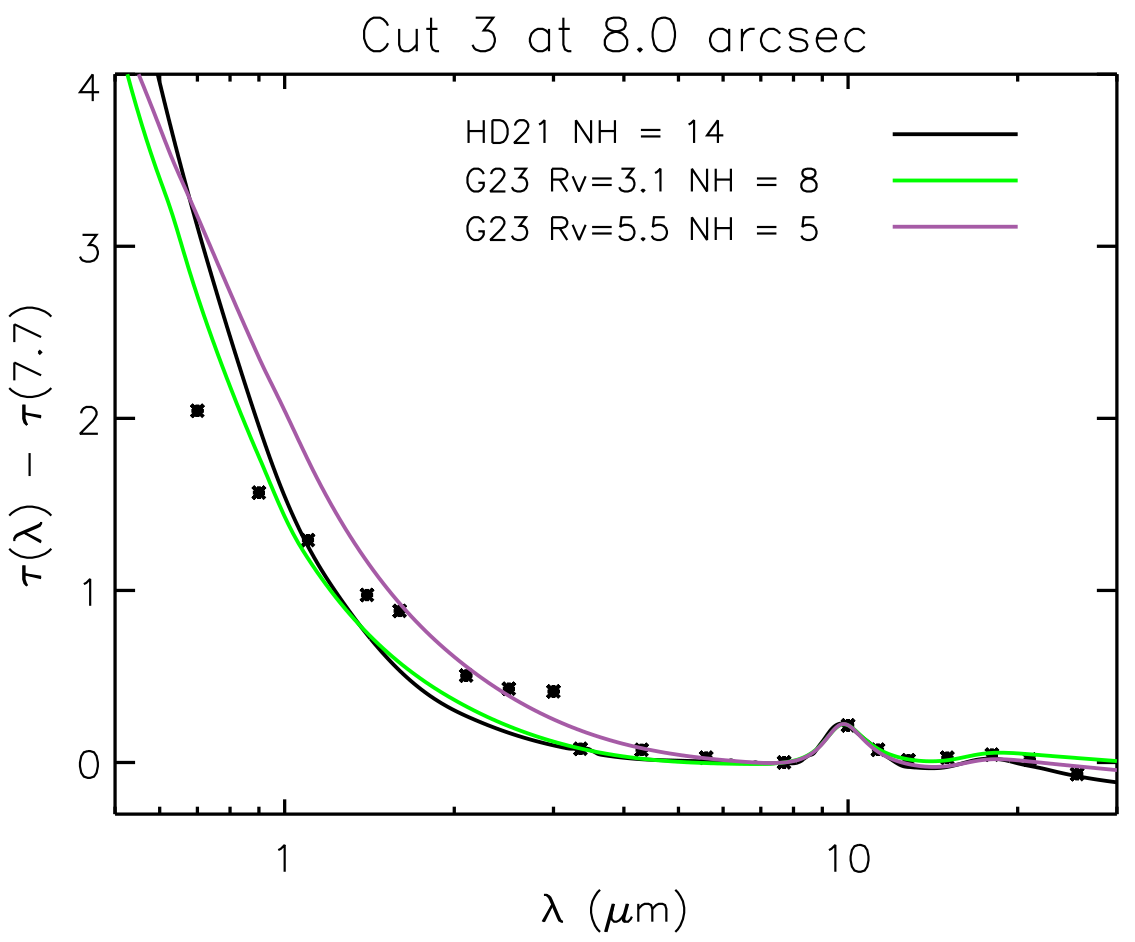

Fig. 14: Spectrum of the optical depth differences $\Delta\tau(\lambda, z_0)$ at the peak position along the third cut shown in Fig. 6. The asterisks are derived from the data. For comparison is shown $\Delta\tau_{ext}(\lambda)$ computed using empirical extinction curves of Gordon et al. (2023) and Hensley & Draine (2021) with the column density (given in units of $10^{21}$ cm$^{-2}$) adjusted so that $\Delta\tau_{ext}(\lambda) = \Delta\tau(\lambda, z_0)$ at 10 $\mu$m.

We also find that the value of $\Delta\tau$ derived from the F300M filter appears to present a small excess for the three cuts. It is likely due to icy $H_2O$ mantles onto grains formed in regions shielded from stellar radiations producing a strong absorbance due to O-H stretch at 3 $\mu$m (e.g., Boogert et al. 2015). This feature is not detectable in the empirical extinction curves measured along diffuse galactic sightlines we use. However it has been predicted in extinction curves seen towards dense sightlines (e.g., Decleir et al. 2022) and simulated for dense regions by Köhler et al. (2015) using dust aggregates with ice mantles.

### 8.3. Discussion

It is interesting to note that our estimated values of the column density necessary to produce the observed attenuation at the peak position of the third cut ($\sim 10^{22}$ H cm$^{-2}$, corresponding to $A_V \sim 5$) is comparable to values deduced by Habart et al. (2005) using 1.2 mm dust observations ($1.5 - 3 \times 10^{22}$ H cm$^{-2}$), and also by Kaplan et al. (2021) using observations of excited rovibrational $H_2$ lines ($A_K = 0.7$, corresponding to $A_V \sim 7$ and $N_H \sim 10^{22}$ H cm$^{-2}$). We see that our very simple model allows us to derive quantitative results consistent with observations made with independent tracers.

The lower left panel of Fig. 12 compares the $\Delta\tau$ profiles with velocity-integrated intensity profiles of the CO $J = 3-2$ and $HCO^+$ $J = 4-3$ lines observed with ALMA by Hernández-Vera et al. (2023) with an angular resolution of 0.5″. The CO $J = 3-2$ profile traces the distribution of the molecular gas, and we see that as expected it increases just behind the illuminated edge. The $HCO^+$ $J = 4-3$ emission traces the dense gas, with density in the range $10^4 - 10^6$cm$^{-3}$ (Goicoechea et al. 2016). The attenuation for the bright IR structures observed at distances of 0 to 5″seems negligible, while the CO emission increases and a faint $HCO^+$ emission is detected. This is likely because the illuminated layer which is observed is located at the extreme edge of the PDR, in the direction of the illuminating star (see Fig. 13), so that the column density of the attenuating molecular matter, detected with the CO and $HCO^+$ lines, should be limited. At greater distances, the attenuation increases, which can be explained by an increase in the column density of molecular matter located between the illuminated layers and the observer traced by the higher $HCO^+$ emission.

As the Horsehead nebula is, in the visible, seen in silhouette against the background, it is not surprising that attenuation effects are significant. To our knowledge, this is the first time that dust attenuation through a PDR has been mapped in such detail. For five nearby PDRs presenting very different excitation and illuminating conditions (Sharpless 140, NGC 2023, IC 63, the Orion Bar and the Horsehead Nebula) Kaplan et al. (2021) obtained $A_K$ values between 0.3 and 0.7 (corresponding to $A_V \sim 3-7$) using observations of excited rovibrational $H_2$ lines. More detailed measurements of rovibrational $H_2$ lines across the Orion Bar with JWST give $A_V$ in a range 3-12 (Peeters et al. 2023; Habart et al. 2023). The dust attenuation seems to be never negligible for PDRs, even in the IR spectral range. The IR galaxies are also affected by dust attenuation due to the large amount of dust present in these galaxies (e.g., Pereira-Santaella et al. 2014). An important point is that the attenuation we are discussing here is not due to the material inside the illuminated layer which emits and scatters radiation, but to material located between this layer and the observer. The attenuation can be non-negligible even if the observed emission or scattering occurs locally at low extinction. This calls for caution in the interpretation of IR observations from nearby PDRs to external galaxies, and in particular in comparison with modeling to derive physical parameters.

## 9. Conclusions

The NIRCam and MIRI imaging observations of the Horsehead nebula allow us to probe its global structure and to resolve its spatial complexity from 0.7 to 25 $\mu$m, with an angular resolution of 0.1 to 1″ (equivalent to $2 \times 10^{-4}$ to $2 \times 10^{-3}$ pc or 40 to 400 au at the Horsehead distance of 400 pc). The unprecedented set of 23 filters we use, combined with two HST filters at 1.1 and 1.6 $\mu$m, allow for the mapping of: (i) the emission of stochastically-heated nano-grains (size below $\sim$ 20 nm), including interstellar PAHs; (ii) the scattering of the incident radiation by large grains; (iii) three lines tracing ionized hydrogen (H$\alpha$, Pa$\alpha$, Br$\alpha$); and (iv) the ro-vibrational 1-0 S(1) line of $H_2$ at 2.12 $\mu$m. This paper presents the first analysis of the observations, and the main results can be summarized as follows:

- We detect a network of faint striated features extending perpendicular to the PDR front into the H II region in the F335M filter (which traces 3.3 $\mu$m/3.4 $\mu$m aromatic and aliphatic features with continuum emitted by nano-grains), in the F770M filter (dominated by the 7.7 $\mu$m aromatic feature), in the F110M HST filter (which traces scattered light by large grains) and in the three filters sensitive to ionized hydrogen (F070W, F187N, and F405N). It may be the first detection of the entrainment of dust particles in the evaporative flow.
- Except for one filament, we do not detect any bright structure with a width of less than 1.5″(corresponding to 3 $10^{-3}$ pc or 600 au) along the illuminated edge of the Horsehead, which corresponds to twice the angular resolution for the F2550W filter. Most of the complexity of the spatial distribution in the NIR and MIR appears to be resolved with JWST.
- The comparison with spectroscopic observations of the H$\alpha$ line with HST-STIS shows that the ionization front is properly resolved in two narrowband NIRCam filters, F187N and F405N, which trace the Pa$\alpha$ and Br$\alpha$ lines, respectively. The measured emissions in these three lines follow the predictions using the case B recombination scaling in the H II region but deviate into the neutral region. This is likely due to differential attenuation in the neutral medium, indicating that the illuminating star $\sigma$-Orionis illuminates the Horsehead from its backside at an oblique angle.
- The filamentary structure of the $H_2$ line at 2.12 $\mu$m at the illuminated edge of the PDR is spatially resolved. It presents numerous sharp sub-structures on scales as small as 1.5″. Compared to maps tracing the nano-grain emission or the large grain scattering, an excess of $H_2$ emission is found all along the edge of the PDR, in a narrow layer (width around 1″, corresponding to 2 $\times 10^{-3}$ pc or 400 au) directly illuminated by $\sigma$-Orionis.
- The ionization front and the dissociation front, detected behind the external edge of the PDR at distances $1-2''$, appear to spatially coincide. This translates into an upper limit of the distance between the two fronts, corresponding to the thickness of the neutral atomic layer, below $\sim$ 100 au. Our observations are compatible with the merging of the two fronts suggested by dynamical models for low-illumination PDRs, such as the Horsehead, but stationary models with high gas pressure cannot be excluded when seeking to explain the observations. The in situ formation of $H_2$ will also have to be taken into account for complete modeling.
- The filamentary structures of the nano-grains emission and of the dust scattering are properly resolved along the edge of the PDR. Our 19 maps taken with broadband filters look quite similar, but strong color variations are found between the illuminated edge and the internal regions. In this first analysis, we show that on a large scale, most of the color variations we observe can be explained by dust attenuation effects – in the scenario where $\sigma$-Orionis is slightly inclined compared to the plan of the sky so that the Horsehead is illuminated from its back side.
- With a very simple model, we are able to derive from our data the main spectral features of the extinction curve, namely, the decrease in the extinction in the NIR spectral range from 0.7 to $\sim$ 5 $\mu$m and extinction excess around 9.7 $\mu$m and 18 $\mu$m. Deviations from the expected extinction curves may be due to to multiple scattering by large grains located inside the PDR or abundance variations of nano-grains. We also find a small excess of extinction at 3 $\mu$m, which may be attributed to icy $H_2O$ mantles onto grains formed in dense regions. For the first time, we derived attenuation profiles across a nearby PDR from 0.7 to 25 $\mu$m. It appears that for lines of sight crossing the inner regions of the Horsehead (and especially around the IR peak of the PDR), extinction effects are non-negligible over the entire spectral range of the JWST.

At this point, a detailed modeling of the dust emission and scattering is necessary in order to disentangle the effects of attenuation and dust evolution across the PDR. Our JWST program also includes IFU spectroscopy with NIRSpec and MIRI, which will be published in future papers. The spectroscopy of the dust emission and scattering ought to set constraints on the physical properties of the dust particles, as done by Elyajouri et al. (2024) for the Orion Bar. The spectroscopy of the $H_2$ lines will enable us to derive the attenuation profiles independently and also to constrain the local physical conditions (temperature, density, and pressure) from the inner regions to the illuminated edge of the PDR, where matter is compressed and undergoes photoevaporation.

*Acknowledgements.* We thank the anonymous referee for very helpful suggestions and comments. NIRCam data reduction is performed at the Steward Observatory. MIRI data reduction is performed at the French MIRI centre of expertise with the support of CNES and the ANR-labcom INCLASS between IAS and the company ACRI-ST. This work is based on observations made with the NASA/ESA/CSA *James Webb* Space Telescope. The data were obtained from the Mikulski Archive for Space Telescopes at the Space Telescope Science Institute, which is operated by the Association of Universities for Research in Astronomy, Inc., under NASA contract NAS 5-03127 for JWST. Part of this work was supported by the Programme National "Physique et Chimie du Milieu Interstellaire" (PCMI) of CNRS/INSU with INC/INP co-funded by CEA and CNES.

[1] Institut d'Astrophysique Spatiale, Université Paris-Saclay, CNRS, 91405 Orsay, France
[2] Steward Observatory, University of Arizona, Tucson, AZ 85721-0065, USA
[3] Space Telescope Science Institute, 3700 San Martin Drive, Baltimore, MD, 21218, USA
[4] Sorbonne Université, CNRS, Institut d'Astrophysique de Paris, 98 bis bd Arago, 75014 Paris, France
[5] Ritter Astrophysical Research Center, University of Toledo, Toledo, OH 43606, USA
[6] Institut de Recherche en Astrophysique et Planétologie, Université Toulouse III - Paul Sabatier, CNRS, CNES, 9 Av. du colonel Roche, 31028 Toulouse, France
[7] Sterrenkundig Observatorium, Universiteit Gent, Gent, Belgium
[8] Max Planck Institute for Astronomy, Königstuhl 17, 69117 Heidelberg, Germany
[9] Université Paris-Saclay, Université Paris Cité, CEA, CNRS, AIM, 91191 Gif-sur-Yvette, France
[10] Leiden Observatory, Leiden University, P.O. Box 9513, 2300 RA Leiden, The Netherlands
[11] Faculty of Aerospace Engineering, Delft University of Technology, Kluyverweg 1, 2629 HS Delft, The Netherlands
[12] European Space Agency, Space Telescope Science Institute, 3700 San Martin Drive, Baltimore, MD, 21218, USA
[13] UK Astronomy Technology Centre, Royal Observatory Edinburgh, Blackford Hill, Edinburgh EH9 3HJ, UK

## Appendix A: Appendix

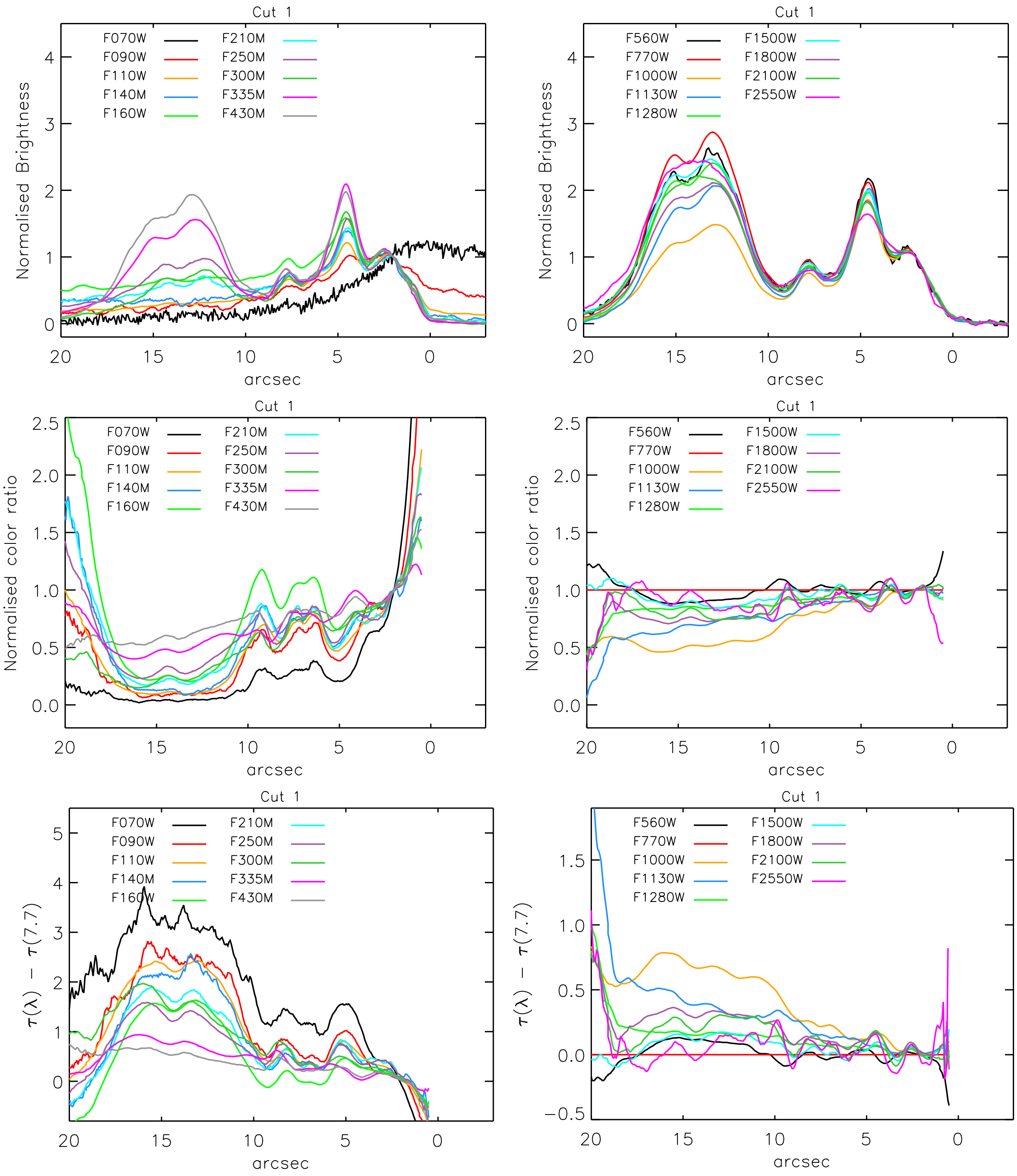

Fig. A.1: Same as Fig. 12 for cut 1

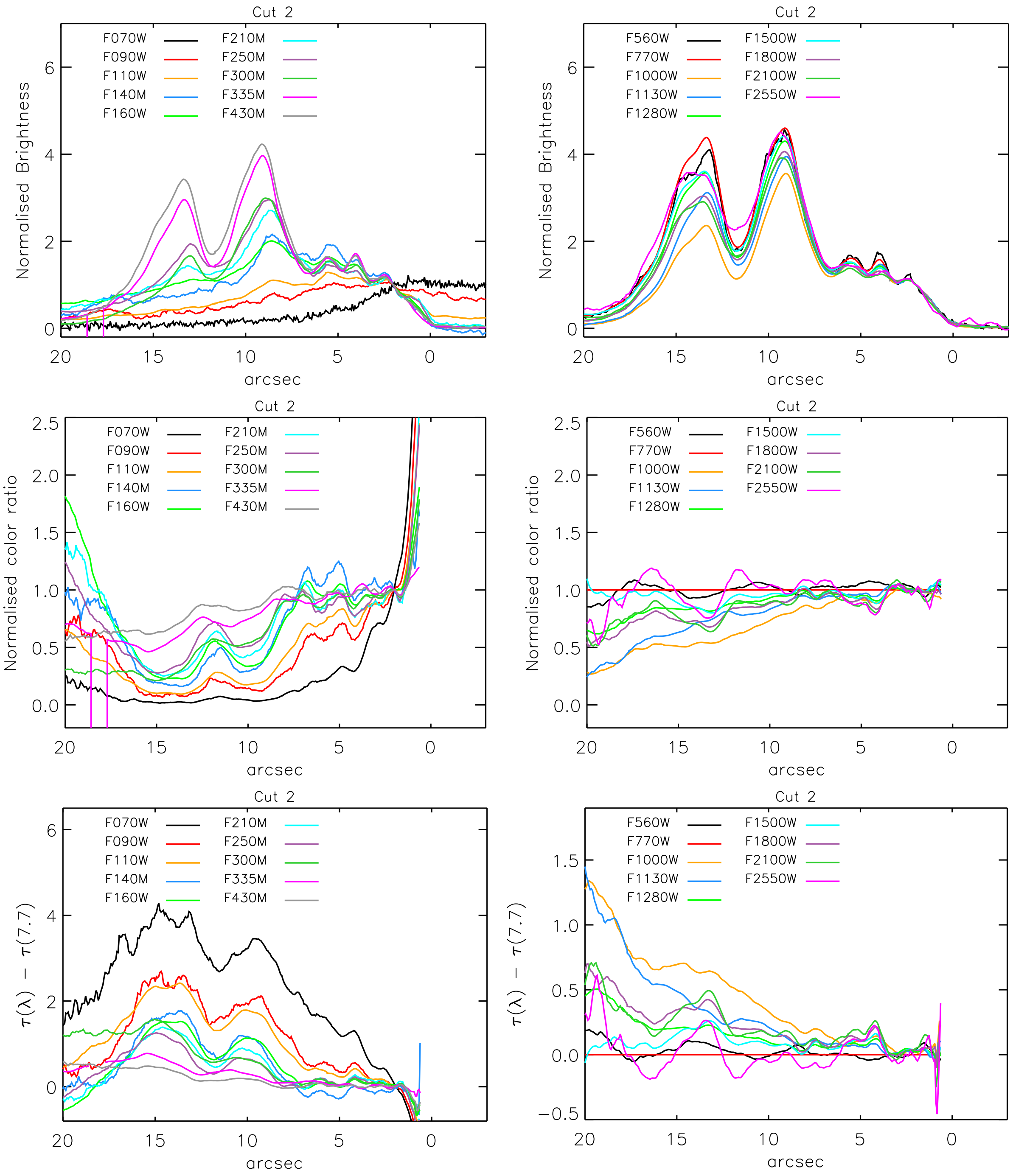

Fig. A.2: Same as Fig. 12 for cut 2

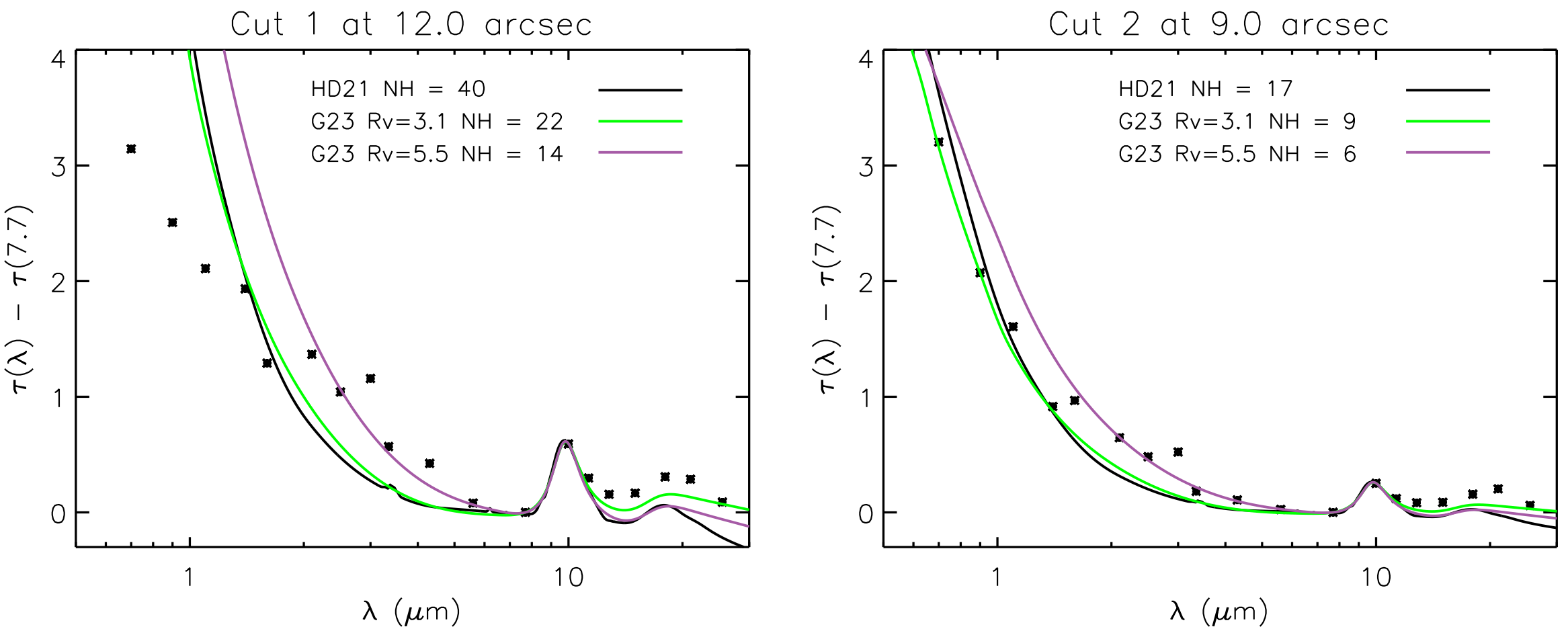

Fig. A.3: Same as Fig. 14 for cuts 1 and 2